\documentclass[aps, prd, showkeys, showpacs ]{revtex4}

\usepackage{amsmath,amsfonts,amscd,amssymb,epsf, multirow}
\usepackage[small]{caption}

\newcommand{\pa}{\partial}

\newcommand{\vep}{\varepsilon}

\begin{document}

\title{Finite temperature  Casimir interaction  between  spheres  in $\boldsymbol{(D+1)}$-dimensional  spacetime: Exact computations and asymptotic expansions}

\author{L. P. Teo}
 \email{LeePeng.Teo@nottingham.edu.my}
 \affiliation{Department of Applied Mathematics, Faculty of Engineering, University of Nottingham Malaysia Campus, Jalan Broga, 43500, Semenyih, Selangor Darul Ehsan, Malaysia.}
\begin{abstract}
We consider the finite temperature Casimir interaction between two Dirichlet spheres in $(D+1)$-dimensional Minkowski spacetime. The Casimir interaction free energy is derived from the zero temperature Casimir interaction energy via the Matsubara formalism. In the high temperature region, the Casimir interaction is dominated by the term with zero Matsubara frequency, and it is known as the classical term since this term is independent of the Planck constant $\hbar$. Explicit expression of the classical term is derived and it is computed exactly using appropriate similarity transforms of matrices. We then compute the small separation asymptotic expansion of this classical term up to the next-to-leading order term. For the remaining part of the finite temperature Casimir interaction with nonzero Matsubara frequencies, we obtain its small separation asymptotic behavior by applying certain prescriptions to the corresponding asymptotic expansion at zero temperature. This gives us a leading term that is shown to agree precisely with the proximity force approximation at any temperature. The next-to-leading order term  at any temperature is also derived and it is expressed as an infinite sum over integrals. To obtain the asymptotic expansion at the low and medium temperature regions, we apply the  inverse Mellin transform techniques. In the low temperature region, we obtain results that agree with our previous work on the zero temperature Casimir interaction.
\end{abstract}

\maketitle
\section{Introduction}

Casimir effect plays an important role in various areas of physics and mathematics such as quantum field theory, gravitation and cosmology, atomic physics, condensed matter, nanotechnology and mathematical physics (see e.g. \cite{9}). From the perspective of quantum field theory, Casimir effect is closely related to the one-loop effective action \cite{14}.  Since the beginning of the last century, physicists have explored higher dimensional spacetime models to solve some fundamental problems such as the unification of fundamental forces and the dark energy and cosmological constant problem. This has motivated the study of Casimir effect in spacetime with arbitrary dimensions \cite{10,11,12,13,15,16,19,17,18,20,21,22}.

Before the turn of this century, the studies of Casimir interaction were concentrated on the parallel plate configuration due to the lack of machineries to compute the Casimir interactions between arbitrary objects. For other configurations such as the sphere-plate configuration which is a popular experimental setup, proximity force approximation was employed to compute the approximation to the Casimir interaction when the separation between the objects is very small compared to the diameter of the objects. However, the situation changed drastically in the beginning of this century. Various approaches have been proposed to compute the Casimir interaction between two objects in (3+1)-dimensional Minkowski spacetime which can either be classified as   worldline method \cite{23,24,25,26,27}, multiple scattering approach \cite{28,29,30,31,32,33,34,35,36,37,38,39,40,41,42} or mode summation approach \cite{43,44,45,46}. Generalizing the prescription in \cite{46} to arbitrary dimensions, we have been able to compute the zero temperature Casimir interaction between a sphere and a plate \cite{47} and between two spheres \cite{1} in $(D+1)$-dimensional Minkowski spacetime.

Motivated by the work \cite{5} which shows that the high temperature limit of the Dirichlet Casimir interaction of the sphere-plate and sphere-sphere configurations can be computed exactly, we have computed the high temperature limit of the Dirichlet sphere-plate interaction in \cite{4}.  In this work, we generalize both the works \cite{1} and \cite{4} and consider the Casimir interaction between two Dirichlet spheres at finite temperature in $(D+1)$-dimensional Minkowski spacetime. In the high temperature limit, we show that the Casimir interaction can also be computed exactly. We then proceed to compute the the leading order term and next-to-leading order terms of the small separation asymptotic expansions of the Casimir interaction at any temperature. Explicit formulas were derived for the low temperature, medium temperature as well as high temperature regions.

We use units where $\hbar=c=k_B=1$.

\section{The finite temperature Casimir interaction between two spheres}
In this work, we consider the finite temperature Casimir interaction between two spheres with radii $R_1$ and $R_2$ respectively. Without loss of generality, we assume that $R_1\leq R_2$ throughout this article. Let   $L$ be the center-to-center distance of the spheres, and let $d=L-R_1-R_2$ be the distance between the two spheres.

In \cite{1}, we showed that when $D\geq 4$, the zero temperature Casimir interaction energy between two Dirichlet spheres with radii $R_1$ and $R_2$   can be  written as
\begin{align}\label{eq12_3_14}
E_{\text{Cas}}^{T= 0}=\frac{1}{2\pi}\int_0^{\infty} d \kappa \sum_{m=0}^{\infty}\frac{(2m+D-3)(m+D-4)!}{(D-3)!m!}\text{Tr}\,\ln\left(1-\mathbb{M}_{m}(\kappa)\right),
\end{align}
where the elements $M_{m;l,l'}$ of the matrix $\mathbb{M}_{m}$ is
\begin{align}
M_{m;l,l'}=&T_{l}^1 \sum_{l''=m}^{\infty}G^1_{m;l,l''}T_{l''}^2G_{m;l'',l'}^2,
\end{align}with
\begin{equation}
\begin{split}
& T_{l}^{i}(\kappa)=\frac{I_{l+\frac{D-2}{2}}(\kappa R_i)}{K_{l+\frac{D-2}{2}}(\kappa  R_i)},
\end{split}
\end{equation}and
 \begin{equation}\begin{split}G^1_{m;l,l'}=G^2_{m;l',l}=&(-1)^{l+m} 2^{2m+D-3}   \Gamma\left(m+\frac{D-2}{2}\right)^2
\sqrt{\frac{\left(l+\frac{D-2}{2}\right)\left(l'+\frac{D-2}{2}\right)(l-m)!(l'-m)!}{(l+m+D-3)!(l'+m+D-3)!}}
\\&\times\int_{0}^{\infty}d\theta
 \left(\sinh\theta\right)^{2m +D-2}
C_{l-m}^{m+\frac{D-2}{2}}\left(\cosh\theta\right)C_{l'-m}^{m+\frac{D-2}{2}}\left(\cosh\theta\right)e^{-\kappa L\cosh\theta}.
\end{split}
\end{equation}
 For fixed $m$, the trace Tr over $\ln\left(1-\mathbb{M}_{m}(\kappa)\right)$ is
\begin{align*}
\sum_{l=m}^{\infty}.
\end{align*}

When $D=3$, we can also represent the Casimir free interaction energy by \eqref{eq12_3_14} provided that the summation $\displaystyle\sum_{m=0}^{\infty}$ is replaced by the summation $\displaystyle\sideset{}{'}\sum_{m=0}^{\infty}$, where the prime $\prime$ indicates that the term $m=0$ is summed with weight $1/2$.

Using Matsubara formalism, the finite temperature Casimir free interaction energy between the two spheres can be obtained by replacing  $ \kappa  $ by the Matsubara frequencies $ \kappa_p= 2\pi p  T$, and replace the integration over $\kappa$ by the summation  over $p$.  Namely,
   \begin{align}\label{eq3_20_1}
E_{\text{Cas}}=T\sideset{}{'}\sum_{p=0}^{\infty}\sum_{m=0}^{\infty}\frac{(2m+D-3)(m+D-4)!}{(D-3)!m!}\text{Tr}\,\ln\left(1-\mathbb{M}_{m}(\kappa_p)\right).
\end{align}

Eq. \eqref{eq3_20_1} can be rewritten as a sum of two terms:
\begin{align*}
E_{\text{Cas}}=E_{\text{Cas}}^{\text{classical}}+E_{\text{Cas}}^{\text{rem}},
\end{align*}
where the classical Casimir energy $E_{\text{Cas}}^{\text{classical}}$ is given by
the $p=0$ term in \eqref{eq3_20_1}:
\begin{align}\label{eq5_22_3}
E_{\text{Cas}}=\frac{T}{2} \sum_{m=0}^{\infty}\frac{(2m+D-3)(m+D-4)!}{(D-3)!m!}\lim_{\kappa\rightarrow 0}\text{Tr}\,\ln\left(1-\mathbb{M}_{m}(\kappa)\right);
\end{align}
and the sum of the remaining terms are denoted by $E_{\text{Cas}}^{\text{rem}}$:
  \begin{align}\label{eq5_22_4}
E_{\text{Cas}}^{\text{rem}}=T\sum_{p=1}^{\infty} \sum_{m=0}^{\infty}\frac{(2m+D-3)(m+D-4)!}{(D-3)!m!}\text{Tr}\,\ln\left(1-\mathbb{M}_{m}(\kappa_p)\right).
\end{align}
The classical term of the Casimir interaction \eqref{eq5_22_3} is the high temperature (i.e., when $1\ll dT$) limit of the free energy.

First, we will compute explicitly the classical term.
As in \cite{4},
we find that as $\kappa\rightarrow 0$,
\begin{equation}
T_{l}^{i}(\kappa)\sim \frac{1}{2^{2l+D-3}\Gamma\left(l+\frac{D-2}{2}\right)\Gamma\left(l+\frac{D}{2}\right)}\left( \kappa R_i \right)^{2l+D-2},
\end{equation}
\begin{equation}
\begin{split}
&\int_{0}^{\infty}d\theta
 \left(\sinh\theta\right)^{2m +D-2}
C_{l-m}^{m+\frac{D-2}{2}}\left(\cosh\theta\right)C_{l'-m}^{m+\frac{D-2}{2}}\left(\cosh\theta\right)e^{-\kappa L\cosh\theta}\\
\sim &\frac{1}{\kappa^{l+l'+D-2}}\frac{2^{l+l'-2m}}{(l-m)!(l'-m)!}
\frac{\Gamma\left(l+\frac{D-2}{2}\right)\Gamma\left(l'+\frac{D-2}{2}\right)}{\Gamma\left(m+\frac{D-2}{2}\right)^2}\frac{\Gamma\left(l+l'+D-2\right)}{L^{l+l'+D-2}}.
\end{split}
\end{equation}
Since the trace of a matrix is not changed if the matrix is replaced by a similar matrix, define
\begin{align}
\widetilde{\mathbb{M}}_m=\mathbb{P}_m^{-1}\mathbb{M}_m\mathbb{P}_m,
\end{align}
where $\mathbb{P}_m$ is a diagonal matrix with elements
\begin{align}
\mathbb{P}_{m;l,l'}=(-1)^l\left(\frac{\kappa L}{2}\right)^{l}\sqrt{\frac{\Gamma(l+m+D-2)}{\left(l+\frac{D-2}{2}\right)(l-m)!}}\frac{1}{\Gamma\left(l+\frac{D-2}{2}\right)}\delta_{l,l'}.
\end{align}
Then
\begin{align}
\widetilde{M}_{m;l,l'}(\kappa) =(-1)^{-l+l'}\left(\frac{\kappa L}{2}\right)^{-l+l'}\sqrt{\frac{\left(l+\frac{D-2}{2}\right)(l-m)!\Gamma(l'+m+D-2)}{\left(l'+\frac{D-2}{2}\right)(l'-m)!\Gamma(l+m+D-2)}}\frac{\Gamma\left(l+\frac{D-2}{2}\right)}
{\Gamma\left(l'+\frac{D-2}{2}\right)}M_{m;l,l'}(\kappa),
\end{align}
and one can deduce that
\begin{align}
\widetilde{M}_{m;l,l'}(0) =&\sum_{l''=m}^{\infty}\frac{(l+l''+D-3)!(l'+l''+D-3)!}{(l+m+D-3)!(l''-m)!(l''+m+D-3)!(l'-m)!}\left(\frac{R_1}{L}\right)^{2l+D-2}\left(\frac{R_2}{L}\right)^{2l''+D-2}.
\end{align}
Let
\begin{align}
b_i=\frac{R_i}{L}.
\end{align}
Then the classical Casimir interaction energy between two spheres in $(D+1)$-dimensional Minkowski spacetime can be written as
\begin{align}\label{eq5_14_1}
E_{\text{Cas}}^{\text{classical}}=\frac{T}{2} \sum_{m=0}^{\infty}\frac{(2m+D-3)(m+D-4)!}{(D-3)!m!}\ln\det\left(\mathbb{I}-\mathbb{N}_m\right),
\end{align}where
\begin{align}\label{eq5_15_1}
N_{m;l,l'}  =&\sum_{l''=m}^{\infty}\frac{(l+l''+D-3)!(l'+l''+D-3)!}{(l+m+D-3)!(l''-m)!(l''+m+D-3)!(l'-m)!}b_1^{2l+D-2}b_2^{2l''+D-2}.
\end{align}

\section{The proximity force approximation}
The proximity force approximation approximates the Casimir interaction force between two objects by summing  the local Casimir force density between two planes over the surfaces.
In $(D+1)$-dimensional Minkowski spacetime, the finite temperature  Casimir   force density between two parallel Dirichlet plates   is given by \cite{6}:
\begin{align}
\mathcal{F}_{\text{Cas}}^{  \parallel} (d)=-\frac{T}{2^{D-3}\pi^{\frac{D-1}{2}}\Gamma\left(\frac{D-1}{2}\right)}\sideset{}{'}\sum_{p=0}^{\infty} \int_{\kappa_p}^{\infty} dx\,\left(x^2-\kappa_p^2\right)^{\frac{D-3}{2}}\frac{x^2}{e^{2dx}-1},
\end{align}where $d$ is the distance between the two plates.

As in \cite{1}, the proximity force approximation to the finite temperature Casimir  interaction force between two Dirichlet spheres in $(D+1)$-dimensional spacetime is
\begin{equation}\label{eq3_26_10}\begin{split}
F_{\text{Cas}}^{ \text{PFA}}  =&R_1^{D-1}\int_0^{\pi}d\theta_1\sin^{D-2}\theta_1\int_0^{\pi}d\theta_2\sin^{D-3}\theta_2\ldots\int_0^{\pi}d\theta_{D-2}\sin\theta_{D-2}\int_{-\pi}^{\pi}
d\theta_{D-1} \mathcal{F}_{\text{Cas}}^{\parallel}\left(d\left(\boldsymbol{\theta}\right)\right)\\
=&\frac{2\pi^{\frac{D-1}{2}}}{\Gamma\left(\frac{D-1}{2}\right)}R_1^{D-1}\int_0^{\pi}d\theta_1 \sin^{D-2}\theta_1 \mathcal{F}_{\text{Cas}}^{\parallel}\left(d\left(\theta_1\right)\right),
\end{split}\end{equation}where
\begin{align}
d\left(\boldsymbol{\theta}\right)=d\left(\theta_1\right)=\sqrt{L^2-2R_1L\cos\theta_1+R_1^2}-R_2
\end{align}is the distance between a point on the surface of the sphere with radius $R_1$ to the sphere with radius $R_2$. Let $u=d\left(\theta_1\right)/d$. Then
\begin{equation}\label{eq5_23_7}\begin{split}
F_{\text{Cas}}^{ \text{PFA}}  =&\frac{2\pi^{\frac{D-1}{2}}}{\Gamma\left(\frac{D-1}{2}\right)}R_1^{D-1}\int_1^{\frac{2R_1+d}{d}}
du \frac{d(du+R_2)}{R_1L}\left(\frac{d(2R_2+du+d)(2R_1+d-du)(2R_1+2R_2+du+d)(u-1)}{(2R_1L)^2}\right)^{\frac{D-3}{2}}\\
&\hspace{3cm}\times  \mathcal{F}_{\text{Cas}}^{\parallel}(du)\\
\sim &\frac{(2\pi)^{\frac{D-1}{2}}}{\Gamma\left(\frac{D-1}{2}\right)}\left(\frac{dR_1R_2}{R_1+R_2}\right)^{\frac{D-1}{2}}\int_1^{\infty} du (u-1)^{\frac{D-3}{2}} \mathcal{F}_{\text{Cas}}^{\parallel}(du)\\
=&-\frac{T}{2^{\frac{D-5}{2}}\Gamma\left(\frac{D-1}{2}\right)^2}\left(\frac{dR_1R_2}{R_1+R_2}\right)^{\frac{D-1}{2}}\sideset{}{'}\sum_{p=0}^{\infty} \int_0^{\infty} du u^{\frac{D-3}{2}} \int_{\kappa_p}^{\infty} dx\,\left(x^2-\kappa_p^2\right)^{\frac{D-3}{2}}\frac{x^2}{e^{2dx(u+1)}-1}\\
=&-\frac{T}{2^{\frac{D-5}{2}}\Gamma\left(\frac{D-1}{2}\right)^2}\left(\frac{ R_1R_2}{R_1+R_2}\right)^{\frac{D-1}{2}}\frac{1}{d^{\frac{D+1}{2}}}\sideset{}{'}\sum_{p=0}^{\infty} \int_0^{\infty} du u^{\frac{D-3}{2}} \int_{\kappa_p d}^{\infty} dx\,\left(x^2-\kappa_p^2d^2\right)^{\frac{D-3}{2}}\frac{x^2}{e^{2x(u+1)}-1}\\
=&-\frac{T}{2^{D-3}\Gamma\left(\frac{D-1}{2}\right)}\left(\frac{ R_1R_2}{R_1+R_2}\right)^{\frac{D-1}{2}}\frac{1}{d^{\frac{D+1}{2}}}\sideset{}{'}\sum_{p=0}^{\infty}\sum_{k=1}^{\infty} \frac{1}{k^{\frac{D-1}{2}}} \int_{\kappa_p d}^{\infty} dx\,\left(x^2-\kappa_p^2d^2\right)^{\frac{D-3}{2}}x^{-\frac{D-5}{2}}e^{-2kx}.
\end{split}\end{equation}
From here, we find that the proximity force approximation to the classical term (i.e., $p=0$ term) of the Casimir interaction force is
\begin{equation}\label{eq5_22_1}\begin{split}
F_{\text{Cas}}^{ \text{classical},\text{PFA}}  =&-\frac{T}{2^{D-2}\Gamma\left(\frac{D-1}{2}\right)}\left(\frac{ R_1R_2}{R_1+R_2}\right)^{\frac{D-1}{2}}\frac{1}{d^{\frac{D+1}{2}}} \sum_{k=1}^{\infty} \frac{1}{k^{\frac{D-1}{2}}} \int_{0}^{\infty} dx\,x^{\frac{D-1}{2}}e^{-2kx}\\
 =&-\frac{(D-1)T\zeta(D)}{2^{\frac{3D-1}{2}} }\left(\frac{ R_1R_2}{R_1+R_2}\right)^{\frac{D-1}{2}}\frac{1}{d^{\frac{D+1}{2}}};
\end{split}\end{equation}
and the proximity force approximation to the zero temperature Casimir interaction force is
\begin{equation}\label{eq5_22_2}\begin{split}
F_{\text{Cas}}^{ T=0,\text{PFA}}  =&-\frac{1}{2^{D-2}\pi\Gamma\left(\frac{D-1}{2}\right)}\left(\frac{ R_1R_2}{R_1+R_2}\right)^{\frac{D-1}{2}}\frac{1}{d^{\frac{D+1}{2}}} \sum_{k=1}^{\infty} \frac{1}{k^{\frac{D-1}{2}}}\int_0^{\infty}d\kappa \int_{\kappa d}^{\infty} dx\,\left(x^2-\kappa^2 d^2\right)^{\frac{D-3}{2}}x^{-\frac{D-5}{2}}e^{-2kx}\\
 =&-\frac{ \Gamma\left(\frac{D+3}{2}\right)\zeta(D+1)}{2^{\frac{3D+1}{2}} \sqrt{\pi}\Gamma\left(\frac{D}{2}\right)}\left(\frac{ R_1R_2}{R_1+R_2}\right)^{\frac{D-1}{2}}\frac{1}{d^{\frac{D+3}{2}}}.
\end{split}\end{equation}

\section{The classical Casimir interaction}\label{classical}

In this section, we derive an alternative expression for the classical Casimir interaction energy \eqref{eq5_14_1} using similarity transforms of matrices. We then compute the asymptotic behavior of the classical Casimir interaction.

Notice that the matrix $\mathbb{N}_m$ \eqref{eq5_15_1} for the sphere-sphere case is more complicated than the corresponding $\mathbb{N}_m$ matrix for the sphere-plate case we obtained in \cite{4}. It can be considered as the multiplication of two different matrices of the type $\mathbb{N}_m$ obtained in \cite{4}. Therefore, the diagonalization of the matrix \eqref{eq5_15_1} is also more complicated.

For fixed $m$, let  $\mathbb{Q}_m$ be a lower triangular matrix with elements
\begin{align}
\left(\mathbb{Q}_m\right)_{l,l'}=\left\{\begin{aligned} \left(\frac{b_1}{b_2}\right)^{-l'}b_1^{2l}\frac{(l-m)!}{(l-l')!(l'-m)!},\hspace{1cm} &l\geq l'\\
0,\hspace{3cm} &l<l'\end{aligned}\right..
\end{align}It is easy to check that its inverse matrix is given by
\begin{align}
\left(\mathbb{Q}_m^{-1}\right)_{l,l'}=\left\{\begin{aligned}(-1)^{l-l'} b_1^{-2l'}\left(\frac{b_1}{b_2}\right)^{l}\frac{(l-m)!}{(l-l')!(l'-m)!},\hspace{1cm} &l\geq l'\\
0,\hspace{3cm} &l<l'\end{aligned}\right..
\end{align}
It follows that
\begin{equation}\begin{split}
\left(\mathbb{Q}_m^{-1}\mathbb{N}_m\mathbb{Q}_m\right)_{l,l'}=& \sum_{l_1=m}^{l}(-1)^{l-l_1} \left(\frac{b_1}{b_2}\right)^{l-l'}\frac{(l-m)!}{(l-l_1)!(l_1-m)!}
\sum_{l_2=m}^{\infty}\sum_{l_3=l'}^{\infty}\frac{(l_1+l_2+D-3)!(l_3+l_2+D-3)!}{(l_1+m+D-3)!(l_2-m)!(l_2+m+D-3)!}\\&\times b_1^{2l_3+D-2}b_2^{2l_2+D-2}\frac{1}{(l_3-l')!(l'-m)!}\\
=&\sum_{l_2=m}^{\infty}\sum_{l_1=m}^{l}(-1)^{l-l_1} \left(\frac{b_1}{b_2}\right)^{l-l'}\frac{(l-m)!}{(l-l_1)!(l_1-m)!}
\frac{(l_1+l_2+D-3)!(l_2+l'+D-3)!}{(l_1+m+D-3)!(l_2-m)!(l_2+m+D-3)!(l'-m)!}\\&\times b_1^{2l'+D-2}b_2^{2l_2+D-2}\frac{1}{\left(1-b_1^2\right)^{l_2+l'+D-2}}.
\end{split}\end{equation}
Notice that
\begin{equation}\begin{split}
&\sum_{l_1=m}^{l}(-1)^{l-l_1}  \frac{(l-m)!}{(l-l_1)!(l_1-m)!}
\frac{(l_1+l_2+D-3)! }{(l_1+m+D-3)!(l_2-m)! }\\=&\text{coefficient of $x^{l-l_1}\cdot x^{l_1+m+D-3}$ in}\,(1-x)^{l-m}\cdot \frac{1}{(1-x)^{l_2-m+1}}\\
=&\left\{\begin{aligned}  \frac{(l_2+m+D-3)!}{(l_2-l)!(l+m+D-3)!},\hspace{1cm} &l_2\geq l\\
0,\hspace{3cm} &l_2<l\end{aligned}\right..
\end{split}\end{equation}
Therefore,
\begin{equation}\begin{split}
\left(\mathbb{Q}_m^{-1}\mathbb{N}_m\mathbb{Q}_m\right)_{l,l'}=&\left(\frac{b_1}{b_2}\right)^{l-l'}\sum_{l_2=l}^{\infty}
\frac{(l_2+l'+D-3)!}{(l_2-l)!(l+m+D-3)!(l'-m)!}b_1^{2l'+D-2}b_2^{2l_2+D-2}\frac{1}{\left(1-b_1^2\right)^{l_2+l'+D-2}}\\
=&\frac{(l+l'+D-3)!}{(l+m+D-3)!(l'-m)!}\beta^{l+l'+D-2},
\end{split}\end{equation}
where
\begin{align}
\beta=\frac{b_1b_2}{1-b_1^2-b_2^2}.
\end{align}
Notice that now the matrix $\mathbb{Q}_m^{-1}\mathbb{N}_m\mathbb{Q}_m$ has a form similar to the matrix $\mathbb{N}_m$ in \cite{4} for the sphere-plate case, with $\beta$ playing the role of $R/(2(R+d))$ in \cite{4}, with $R$ the radius of the sphere and $d$ the distance from the sphere to the plane.
In fact, in the limit the radius of the second sphere goes to infinity (i.e., the sphere-plane limit), we have
\begin{align}
\beta=\frac{R_1R_2}{L^2-R_1^2-R_2^2}=\frac{R_1R_2}{(R_1+R_2+d)^2-R_1^2-R_2^2}\xrightarrow{R_2\rightarrow\infty}\frac{R_1}{2(R_1+d)}.
\end{align}

As in \cite{4}, we then find that
 the classical Casimir interaction energy between two Dirichlet spheres is
\begin{equation}\begin{split}\label{eq5_15_2}
E_{\text{Cas}}^{\text{classical}}=&\frac{T}{2}\sum_{l=0}^{\infty}\frac{(2l+D-2)(l+D-3)!}{(D-2)!l!}\ln\left(1-y^{2l+D-2}\right),
\end{split}\end{equation}
where $0<y<1$ is such that
\begin{equation}\label{eq3_26_4}y+y^{-1}=\frac{1}{\beta}.\end{equation}
When $D=3$, we find that
\begin{align}\label{eq5_15_2}
E_{\text{Cas}}^{\text{classical}}=&\frac{T}{2}\sum_{l=0}^{\infty}(2l+1)\ln\left(1-y^{2l+1}\right).
\end{align}
Let us compare this to the result obtained in \cite{5} using bi-spherical coordinates.
In \cite{5}, it is proved that the classical Casimir interaction energy between two Dirichlet spheres in $(3+1)$-dimensional Minkowski spacetime is
\begin{equation}\begin{split}
E_{\text{Cas}}^{\text{classical}}=&\frac{T}{2}\sum_{l=0}^{\infty}(2l+1)\ln\left(1-Z^{2l+1}\right),
\end{split}\end{equation}
where
\begin{equation}\begin{split}
Z=&\frac{1}{\left(\lambda_1+\sqrt{\lambda_1^2-1}\right)\left(\lambda_2+\sqrt{\lambda_2^2-1}\right)},\end{split}\end{equation}
with
\begin{align}
\lambda_1=\frac{1+b_1^2-b_2^2}{2b_1},\hspace{1cm}\lambda_2=\frac{1+b_2^2-b_1^2}{2b_2}.
\end{align}
Hence,
\begin{equation}\begin{split}
Z+Z^{-1}=&\left(\lambda_1-\sqrt{\lambda_1^2-1}\right)\left(\lambda_2-\sqrt{\lambda_2^2-1}\right)+\left(\lambda_1+\sqrt{\lambda_1^2-1}\right)\left(\lambda_2+\sqrt{\lambda_2^2-1}\right)\\
=&2\left(\lambda_1\lambda_2+\sqrt{(\lambda_1^2-1)(\lambda_2^2-1)}\right)\\
=&\frac{1-b_1^2-b_2^2}{b_1b_2}.
\end{split}\end{equation}Since $0<Z<1$, $Z$ coincides with the $y$ defined by \eqref{eq3_26_4}. This shows that the result \eqref{eq5_15_2} we obtain here agrees with the result obtained in \cite{5}. But here we only use the similarity transform of matrices, which is much simpler than using bi-spherical coordinates in \cite{5}.

Next we consider the small separation asymptotic expansion.
Let
\begin{align}
\vep=\frac{d}{R_1+R_2},\hspace{1cm} a_1=\frac{R_1}{R_1+R_2},\hspace{1cm}a_2=\frac{R_2}{R_1+R_2}.
\end{align}Then $a_1+a_2=1$ and
\begin{align}
\frac{1}{\beta}=\frac{(1+\vep)^2-a_1^2-a_2^2}{a_1a_2}.
\end{align}
Let
\begin{equation}\label{eq5_15_4}\begin{split}
\mu=&-\ln y=\ln\left(\frac{1}{2\beta}+\sqrt{\frac{1}{4\beta^2}-1}\right) \\
=&\ln\frac{(1+\vep)^2-a_1^2-a_2^2+\sqrt{(\vep^2+2\vep)(2a_1+\vep)(2a_2+\vep)}}{2a_1a_2}.
\end{split}\end{equation}
Then
\begin{align}
\eta:
=\frac{\pa\mu}{\pa d}=&\frac{2(1+\vep)}{\sqrt{(2a_1+\vep)(2a_2+\vep)(\vep^2+2\vep)}}\frac{1}{R_1+R_2}.
\end{align}
On the other hand, we have an expansion of the form (see \cite{4}):
\begin{align}
\frac{(2l+D-2)(l+D-3)!}{(D-2)!l!}=\frac{2}{(D-2)!}\sum_{j=1}^{D-2}x_{D;j}\left(l+\frac{D-2}{2}\right)^j,
\end{align}where the coefficient  $x_{D;j}$ is nonzero only if $D$ and $j$ are both odd or both even.

As in \cite{4}, we find that when $D$ is even, the classical  Casimir interaction force is given by
\begin{equation}\label{eq3_26_8}\begin{split}
F_{\text{Cas}}^{\text{classical}}=-  \frac{2\eta T}{(D-2)!} \sum_{j=1}^{D-2}x_{D;j}\left\{\frac{\Gamma(j+2)}{2^{j+2}\mu^{j+2}}\zeta(j+2)+(-1)^{\frac{j}{2}}\frac{\Gamma(j+2)}{2^{j+2}\pi^{j+2}}\zeta(j+2)
+\frac{(-1)^{\frac{j}{2}+1}\pi^{j+2}}{\mu^{j+2}}\sum_{n=1}^{\infty} \frac{n^{j+1}}{e^{\frac{2\pi^2 n}{\mu}}-1}\right\};
\end{split}\end{equation}and when $D$ is odd,
\begin{equation}\label{eq3_26_9}\begin{split}
F_{\text{Cas}}^{\text{classical}}=- \frac{2\eta T}{(D-2)!} \sum_{j=1}^{D-2}x_{D;j}\left\{\frac{\Gamma(j+2)}{2^{j+2}\mu^{j+2}}\zeta(j+2)+\frac{(-1)^{\frac{j-1}{2}}}{2^{j+2}}\int_0^{\infty}\frac{y^{j+1}\cot\frac{\mu y}{2}}{e^{\pi y}+1}dy\right\}.
\end{split}\end{equation}
Here $\zeta(s)=\sum_{n=1}^{\infty}n^{-s}$ is the Riemann zeta function.

In principal, one can derive the small separation asymptotic behavior of the classical Casimir interaction force from \eqref{eq3_26_8} and \eqref{eq3_26_9} up to any order in $\vep$. In the following, we only compute the leading order term and the next-to-leading order term.

As $\vep\ll 1$, one can show that
\begin{align}\label{eq3_26_11}
 \eta=&\frac{1}{R_1+R_2}\frac{1}{\sqrt{2\vep a_1 a_2}}\left(1+ \frac{3a_1a_2-1}{4 a_1a_2} \vep+\ldots\right),
\end{align}
\begin{align}\label{eq3_26_12}
\mu=& \sqrt{\frac{2\vep}{a_1a_2}} \left(1+\frac{3a_1a_2-1}{12a_1a_2}\vep+\ldots\right).
\end{align}
For the constants $x_{D;j}$,
\begin{equation}\label{eq3_27_11}\begin{split}
x_{D;D-2}=&1,\quad x_{D;D-3}=0,\quad  x_{D;D-4}=-\frac{(D-2)(D-3)(D-4)}{24},\ldots.
\end{split}\end{equation}
On the other hand,
\begin{equation}\label{eq3_26_13}\begin{split}
\frac{(-1)^{\frac{j-1}{2}}}{2^{j+2}}\int_0^{\infty}\frac{y^{j+1}\cot\frac{\mu y}{2}}{e^{\pi y}+1}dy
=\frac{(-1)^{\frac{j-1}{2}}}{2^{j+1}}\frac{1}{\mu}\frac{\Gamma(j+1)}{\pi^{j+1}}(1-2^{-j})\zeta(j+1)
+O(\mu).
\end{split}\end{equation}
When $D=3$, \eqref{eq3_26_9}, \eqref{eq3_26_13}, \eqref{eq3_26_11} and \eqref{eq3_26_12} give
\begin{equation}\label{eq3_27_5}\begin{split}
F_{\text{Cas}}^{\text{classical}}
=&-\frac{TR_1R_2}{8(R_1+R_2)d^2}\zeta(3)\left\{1+\frac{1}{6 \zeta(3)}\left(\frac{d}{R_1}+\frac{d}{R_2}\right) +\ldots\right\}.
\end{split}\end{equation}

\begin{figure}[h]
\epsfxsize=0.5\linewidth \epsffile{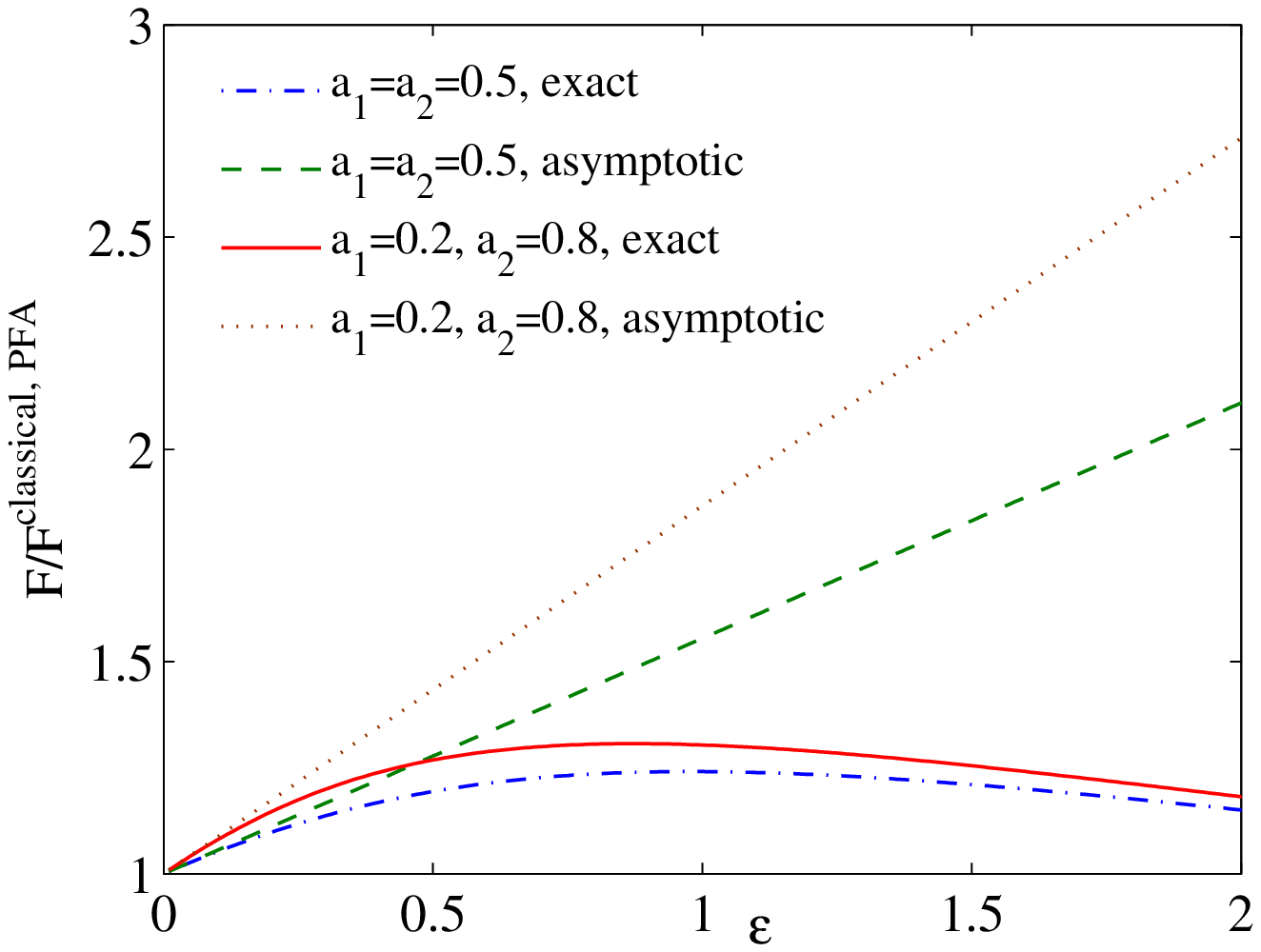} \caption{\label{f1} The comparison between the exact classical Casimir interaction force with the   asymptotic expansion \eqref{eq3_27_5} when $D=3$. Both quantities are normalized by the proximity force approximation. }\end{figure}

\begin{figure}[h]
\epsfxsize=0.5\linewidth \epsffile{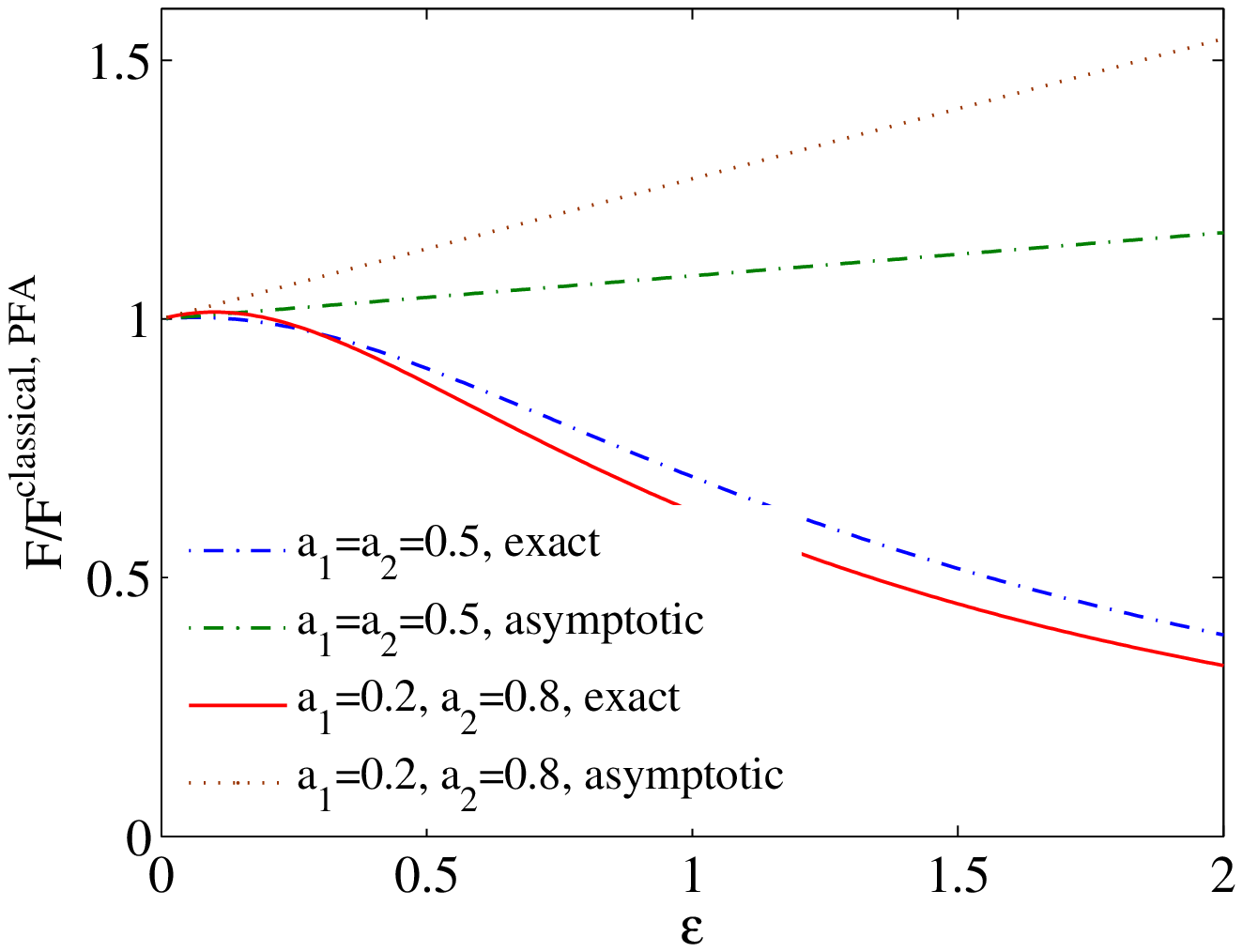} \caption{\label{f2} The comparison between the exact classical Casimir interaction force with the   asymptotic expansion \eqref{eq3_26_14} when $D=4$. Both quantities are normalized by the proximity force approximation. }\end{figure}

When $D\geq 4$, we do not need to take into account the term \eqref{eq3_26_13} in \eqref{eq3_26_9} nor the second term and third term in \eqref{eq3_26_8}. Eqs. \eqref{eq3_26_8}, \eqref{eq3_26_9},  \eqref{eq3_26_11}, \eqref{eq3_26_12}  and  \eqref{eq3_27_11} give
\begin{equation}\label{eq3_26_14}\begin{split}
F_{\text{Cas}}^{\text{classical}}=&-\frac{(D-1)T\zeta(D)}{2^{\frac{3D-1}{2}} }\left(\frac{ R_1R_2}{R_1+R_2}\right)^{\frac{D-1}{2}}\frac{1}{d^{\frac{D+1}{2}}}
\\&\hspace{1cm}\times\left\{1-\frac{D-3}{4}\frac{d}{R_1+R_2}+\left[\frac{D-3}{12 } -\frac{(D-3)(D-4)}{3(D-1) }\frac{\zeta(D-2)}{\zeta(D)}\right]\left(\frac{d}{R_1}+\frac{d}{R_2}\right)+\ldots\right\}.
\end{split}\end{equation}
Notice that the leading term is
\begin{align}\label{eq5_26_2}
F_{\text{Cas}}^{\text{classical}, 0}= -\frac{(D-1)T\zeta(D)}{2^{\frac{3D-1}{2}} }\left(\frac{ R_1R_2}{R_1+R_2}\right)^{\frac{D-1}{2}}\frac{1}{d^{\frac{D+1}{2}}},
\end{align}and it agrees with the proximity force approximation \eqref{eq5_22_1}.
When $D=3$, the next-to-leading order term of the Casimir interaction force is
\begin{align}\label{eq5_30_6}
F_{\text{Cas}}^{\text{classical}, 1}=-\frac{T}{48d};
\end{align}and when $D\geq 4$,
\begin{equation}\label{eq5_26_3}\begin{split}
F_{\text{Cas}}^{\text{classical}, 1}=&-\frac{(D-1)T\zeta(D)}{2^{\frac{3D-1}{2}} }\left(\frac{ R_1R_2}{R_1+R_2}\right)^{\frac{D-1}{2}}\frac{1}{d^{\frac{D+1}{2}}}\\&\times\left\{ -\frac{D-3}{4}\frac{d}{R_1+R_2}+\left[\frac{D-3}{12 } -\frac{(D-3)(D-4)}{3(D-1) }\frac{\zeta(D-2)}{\zeta(D)}\right]\left(\frac{d}{R_1}+\frac{d}{R_2}\right) \right\}.
\end{split}\end{equation} 

\begin{figure}[h]
\epsfxsize=0.5\linewidth \epsffile{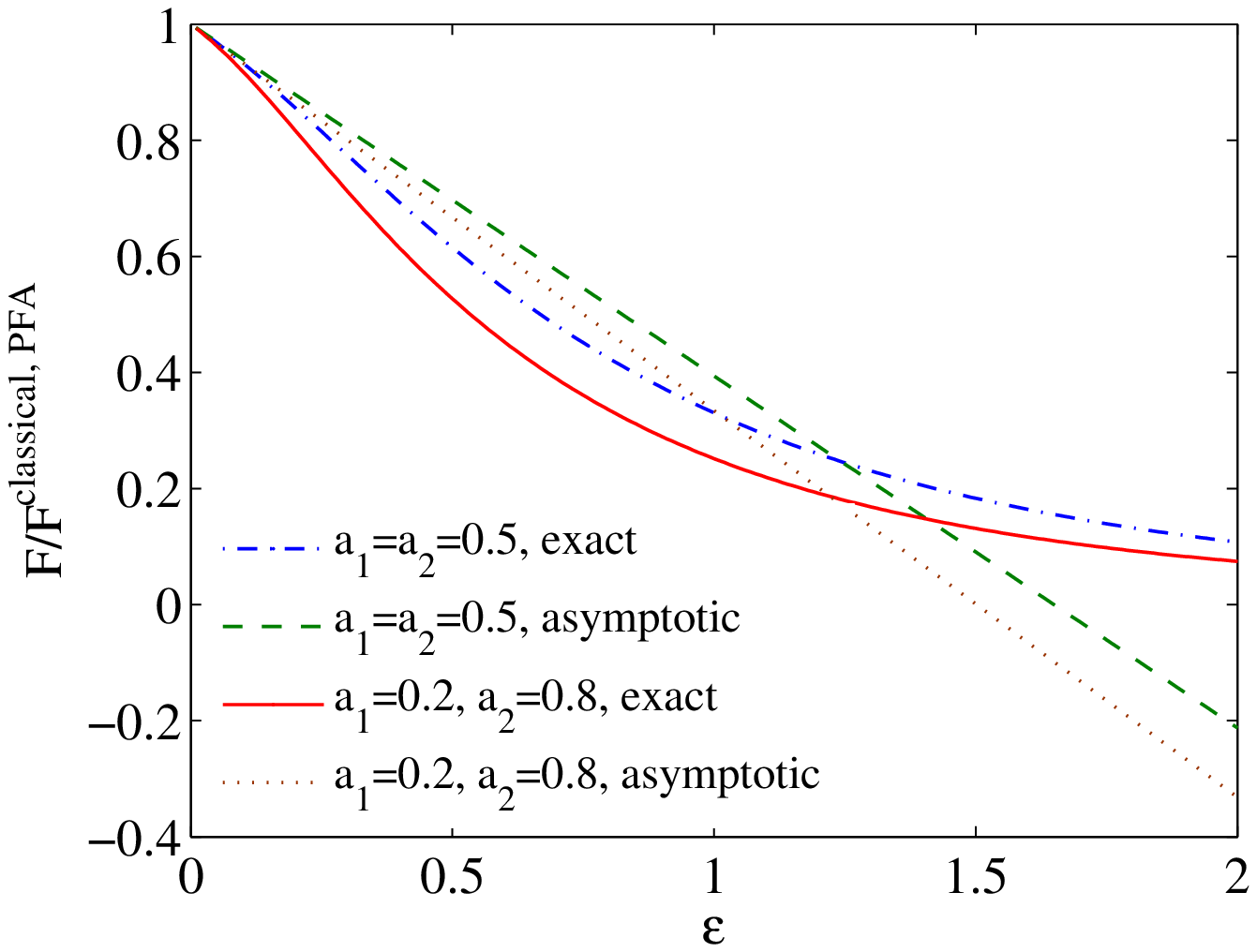} \caption{\label{f3} The comparison between the exact classical Casimir interaction force with the   asymptotic expansion \eqref{eq3_26_14} when $D=5$. Both quantities are normalized by the proximity force approximation. }\end{figure}

\begin{figure}[h]
\epsfxsize=0.5\linewidth \epsffile{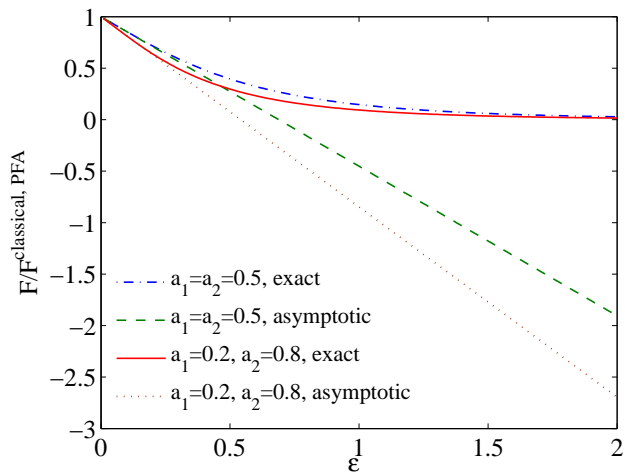} \caption{\label{f4} The comparison between the exact classical Casimir interaction force with the   asymptotic expansion \eqref{eq3_26_14} when $D=6$. Both quantities are normalized by the proximity force approximation. }\end{figure}

In Figs. \ref{f1}, \ref{f2}, \ref{f3} and \ref{f4}, we compare the exact classical Casimir interaction force to the two-term asymptotic expansions   \eqref{eq3_27_5} and \eqref{eq3_26_14} when $D=3, 4, 5$ and $6$.  Both quantities are normalized by the proximity force approximation.

\begin{figure}[h]
\epsfxsize=0.5\linewidth \epsffile{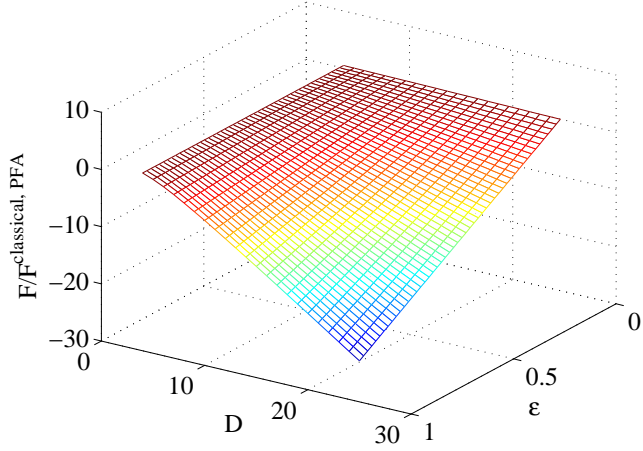} \caption{\label{f5} Dependence of the  asymptotic expansion \eqref{eq3_26_14} on $\vep$ and $D$ for $0\leq \vep\leq 1$ and   $6\leq D\leq 25$. The asymptotic expansion is normalized by the proximity force approximation. }\end{figure}

In Fig. \ref{f5}, we plot the dependence of the two term asymptotic expansion \eqref{eq3_26_14}, normalized by the proximity force approximation, on the normalized distance $\vep$ and dimension $D$. It is observed that the correction to the proximity force approximation becomes larger when $D$ is larger.
In fact, from \eqref{eq3_26_14}, we find that when $D$ is large,
\begin{equation}\begin{split}
F_{\text{Cas}}^{\text{classical}}\sim &F_{\text{Cas}}^{\text{classical, PFA}}\left\{1-\frac{D}{4}\left(\frac{d}{R_1+R_2}+\frac{d}{R_1}+\frac{d}{R_2}\right)+ \ldots\right\}.
\end{split}
\end{equation}
Hence, in the high temperature region, we find that the next-to-leading order term is proportional to $D$, which shows a larger deviation from the proximity force approximation when the dimension $D$ becomes larger.

\section{Small separation asymptotic behavior of the Casimir interaction}

The small separation asymptotic expansion of the Casimir interaction is in general not easy to compute. One usually expects that the leading term would coincide with the proximity force approximation. However, the computation of the next-to-leading order term is often tedious, but it captures important information of the Casimir interaction such as the response of the system to the curvature of the surfaces, and it also gives a rough idea of how accurate the leading term is.   Using the idea first developed in \cite{37}, such computations have been performed for various geometric configurations \cite{37, 38, 48,49,50,51,52,53,54,7,55}.

In this section, we will compute the small separation asymptotic expansions of the Casimir interaction at any temperature.  Since the classical term \eqref{eq5_22_3} has been considered in Section \ref{classical}, we first consider the remaining terms \eqref{eq5_22_4}, and then combine with the results obtained in Section \ref{classical} to obtain the asymptotic expansion of the full interaction.

The small separation asymptotic expansion of the term \eqref{eq5_22_4} can be computed in the same way as in \cite{1} for the zero temperature case.
In fact, since \eqref{eq5_22_4} is obtained from the zero temperature Casimir interaction energy \eqref{eq12_3_14} by replacing $\kappa$ by $\kappa_p$, and changing the integration over $\kappa$ to summation over $p$, one can use this recipe to obtain  the small $\vep$ asymptotic expansion of \eqref{eq5_22_4} from the small $\vep$ asymptotic expansion of the zero temperature Casimir interaction derived in \cite{1}.

In \cite{1}, we have shown that when $\vep\ll 1$, the leading term of the zero temperature Casimir interaction energy $E_{\text{Cas}}^{T=0, 0}$ is given by
\begin{equation}\begin{split}
E_{\text{Cas}}^{T=0, 0}= & -\frac{  a_2^{\frac{D-1}{2}}}{2^{D-1}\pi R_1 \Gamma\left(\frac{D-1}{2}\right)}\sum_{s=0}^{\infty}\frac{1}{\left(s+1\right)^{\frac{D+1}{2}}}
 \int_0^{\infty} dl \, l^{\frac{D-1}{2}} \int_0^{1} \frac{d\tau\,\tau^{\frac{D-5}{2}} }{\sqrt{1-\tau^2}}\exp\left(
 -\frac{2l(s+1)\vep}{a_1\tau} \right);
\end{split}\end{equation}whereas the next-to-leading order term $E_{\text{Cas}}^{T=0, 1}$ can be written as a sum of two terms:
\begin{align}
E_{\text{Cas}}^{T=0, 1}=E_{\text{Cas}}^{T=0, 1a}+E_{\text{Cas}}^{T=0, 1b}.
\end{align} The first term $E_{\text{Cas}}^{T=0, 1a}$ vanishes for $D=3,4,5$, and for $D\geq 6$,
\begin{equation}\begin{split}
 E_{\text{Cas}}^{T=0, 1a}=& \frac{(D-3)(D-5)}{3}\frac{  a_2^{\frac{D-3}{2}}}{ 2^{D+1}\pi R_1  \Gamma\left(\frac{D-1}{2}\right)}\sum_{s=0}^{\infty}\frac{1}{\left(s+1\right)^{\frac{D-1}{2}}}
 \int_0^{\infty} dl \, l^{\frac{D-3}{2}} \int_0^{1} \frac{d\tau\,\tau^{\frac{D-7}{2}} }{\sqrt{1-\tau^2}}\exp\left(
 -\frac{2l(s+1)\vep}{a_1\tau} \right).
\end{split}\end{equation}The second term $E_{\text{Cas}}^{T=0, 1b}=0$ is
\begin{equation} \begin{split}
E_{\text{Cas}}^{T=0, 1b}=  &-\frac{  a_2^{\frac{D-1}{2}}}{2^{D-1}\pi  R_1   \Gamma\left(\frac{D-1}{2}\right)}\sum_{s=0}^{\infty}\frac{1}{\left(s+1\right)^{\frac{D+1}{2}}}
 \int_0^{\infty} dl \, l^{\frac{D-1}{2}}\int_0^{1} \frac{d\tau\, \tau^{\frac{D-5}{2}}}{ \sqrt{1-\tau^2}} \exp\left(
 -\frac{2l(s+1)\vep}{a_1\tau} \right)    \mathcal{A},\end{split}
\end{equation}where
\begin{equation}
\begin{split}
\mathcal{A}=&-\frac{(D-1)\left(5D-7-(9D-15)a_1+(3D-9)a_1^2\right)}{48a_2l(s+1)}\tau-\frac{(D-1)(D+1)(3a_1-2)}{48a_2l(s+1)}\tau^3\\
&+\frac{(D-2)^2}{2l}-\frac{(D-1)(D-2)}{4l}\tau^2+\frac{(D-3)(D-5)}{12a_2l}\frac{(s+1)}{\tau}-\frac{(D-1)(D-5)\tau}{12a_2l}(s+1)+\frac{(D^2-25)\tau^3}{48a_2l} (s+1)\\
&+\frac{\vep(D-1)(3a_1^2-1)}{6a_1a_2}-\frac{\vep(D-1)(3a_1-2)}{6a_1a_2}\tau^2-\frac{\vep(D-2)}{a_1}(s+1)\tau-\frac{\vep(D-1)}{3a_1a_2}(s+1)^2+\frac{\vep(D-1)}{6a_1a_2}(s+1)^2\tau^2\\
&-\frac{\vep^2l(3a_1-2)}{3a_1^2a_2}(s+1)\tau+\frac{\vep^2l}{3a_1^2a_2}(s+1)^3\tau.
\end{split}
\end{equation}
The integration over $\kappa$ has been changed to the integration over $\tau$ where
\begin{align*}
\kappa=\frac{l\sqrt{1-\tau^2}}{R_1\tau}.
\end{align*}
It follows that
\begin{align*}
d\kappa=-\frac{l}{R_1}\frac{d\tau}{\tau^2\sqrt{1-\tau^2}}.
\end{align*}
To obtain the corresponding small separation asymptotic expansion for $E_{\text{Cas}}^{\text{rem}}$ \eqref{eq5_22_4}, we change $\tau$ to
\begin{align*}
\frac{l}{\sqrt{l^2+(R_1\kappa_p)^2}},
\end{align*}
and make the replacement
\begin{align}
\frac{l}{R_1}\int_0^1\frac{d\tau}{\tau^2\sqrt{1-\tau^2}}\mapsto 2\pi T\sum_{p=1}^{\infty}.
\end{align}
Let
$$z=\frac{\vep l}{a_1},\quad k=s+1.$$
With the prescription described above, we find that for the Casimir force
\begin{align}
F_{\text{Cas}}=-\frac{\pa E_{\text{Cas}}}{\pa d}=F_{\text{Cas}}^{\text{classical}}+F_{\text{Cas}}^{\text{rem}},
\end{align} the leading term of $F_{\text{Cas}}^{\text{rem}}$ \eqref{eq5_22_4} is given by
\begin{equation}\label{eq5_23_6}\begin{split}
F_{\text{Cas}}^{\text{rem}, 0}= & -\frac{ T  }{2^{D-3}    \Gamma\left(\frac{D-1}{2}\right)}\left(\frac{R_1R_2}{R_1+R_2}\right)^{\frac{D-1}{2}}\frac{1}{d^{\frac{D+1}{2}}}\sum_{k=1}^{\infty}\frac{1}{k^{\frac{D-1}{2}}}\sum_{p=1}^{\infty}
 \int_0^{\infty} dz \,\frac{ z^{D-2}}{\left(z^2+( \kappa_p d)^2\right)^{\frac{D-3}{4}}} \exp\left(
 - 2 k \sqrt{z^2+( \kappa_p d)^2} \right).
\end{split}\end{equation}
The next-to-leading order term $F_{\text{Cas}}^{\text{rem}, 1}$ can be written as a sum of two terms:
\begin{align}
F_{\text{Cas}}^{\text{rem}, 1}=F_{\text{Cas}}^{\text{rem}, 1a}+F_{\text{Cas}}^{\text{rem}, 1b}.
\end{align}The first term $F_{\text{Cas}}^{\text{rem}, 1a}$ vanishes for $D=3, 4, 5$, and for $D\geq 6$,
\begin{equation}\label{eq5_23_8}\begin{split}
F_{\text{Cas}}^{\text{rem}, 1a}= &  \frac{(D-3)(D-5)}{3}\frac{ T  }{2^{D-1}    \Gamma\left(\frac{D-1}{2}\right)}\left(\frac{R_1R_2}{R_1+R_2}\right)^{\frac{D-3}{2}}\frac{1}{d^{\frac{D-1}{2}}}\\&\times\sum_{k=1}^{\infty}\frac{1}{k^{\frac{D-3}{2}}}\sum_{p=1}^{\infty}
 \int_0^{\infty} dz \,\frac{ z^{D-4}}{\left(z^2+(\kappa_p d)^2\right)^{\frac{D-5}{4}}} \exp\left(
 -  2 k \sqrt{z^2+( \kappa_p d)^2} \right).
\end{split}\end{equation}The second term $F_{\text{Cas}}^{\text{rem}, 1b}$ is
\begin{equation}\label{eq5_26_1}\begin{split}
F_{\text{Cas}}^{\text{rem}, 1b}= & -\frac{ T  }{2^{D-3}    \Gamma\left(\frac{D-1}{2}\right)}\frac{(R_1R_2)^{\frac{D-1}{2}}}{(R_1+R_2)^{\frac{D+1}{2}}} \frac{1}{d^{\frac{D-1}{2}}}\sum_{k=1}^{\infty}\frac{1}{k^{\frac{D-1}{2}}}\sum_{p=1}^{\infty}
 \int_0^{\infty} dz \,\frac{ z^{D-2}}{\left(z^2+( \kappa_p d)^2\right)^{\frac{D-3}{4}}} \exp\left(
 - 2 k \sqrt{z^2+( \kappa_p d)^2} \right)\mathcal{B},
\end{split}\end{equation}
where
\begin{equation}\label{eq5_23_9}\begin{split}
\mathcal{B}=&\frac{(D-3)(D-5)k}{12a_1a_2z}\frac{\sqrt{z^2+(\kappa_p d)^2}}{z}+\frac{(D-2)^2}{2a_1z}+\frac{(D-1)(3a_1^2-1)}{6a_1a_2}-\frac{(D-1)k^2}{3a_1a_2}\\
&+\left\{-\frac{(D-1)\left[5D-11-(9D-15)a_1+(3D+3)a_1^2\right]}{48a_1a_2zk}-\frac{(D-1)(D-7)k }{12a_1a_2z}-\frac{(D-2)k}{a_1}\right.\\&\left.
\hspace{2cm}-\frac{ z(3a_1-2)k}{3a_1a_2} +\frac{k^3 z}{3a_1a_2}\right\}\frac{z}{\sqrt{z^2+(\kappa_p d)^2}}\\
&+\left\{-\frac{(D-3)(D-2)}{4a_1z}-\frac{(D-3)(3a_1-2)}{6a_1a_2}+\frac{(D-3)k^2}{6a_1a_2}\right\}\frac{z^2}{z^2+(\kappa_pd)^2}\\
&+\left\{-\frac{(D-1)(D-3)(3a_1-2)}{48a_1a_2zk}+\frac{(D+3)(D-7)k }{48a_1a_2z} \right\}\frac{z^3}{\left(z^2+(\kappa_pd)^2\right)^{\frac{3}{2}}}.
\end{split}\end{equation}Using integration by parts, one can show that the $\mathcal{B}$ in \eqref{eq5_26_1} can be replaced by
\begin{align*}
\mathcal{B}'=&-\frac{(D-1)(3a_1a_2-1)}{6a_1a_2}+\frac{(D-1)(D+1)(3a_1a_2-1)}{48a_1a_2k}\frac{1}{\sqrt{z^2+(\kappa_pd)^2}}+\frac{(D-3)(D-5)k}{12a_1a_2 }\frac{\sqrt{z^2+(\kappa_p d)^2}}{z^2}
\\&-\frac{(D-1)(D-7)k}{12a_1a_2}\frac{1}{\sqrt{z^2+(\kappa_pd)^2}}+\frac{(D+3)(D-7)k}{48a_1a_2}\frac{z^2}{\left(z^2+(\kappa_pd)^2\right)^{\frac{3}{2}}}
-\frac{k^3}{3a_1a_2}\frac{z^2}{\sqrt{z^2+(\kappa_pd)^2}}.
\end{align*}

\begin{figure}[h]
\epsfxsize=0.5\linewidth \epsffile{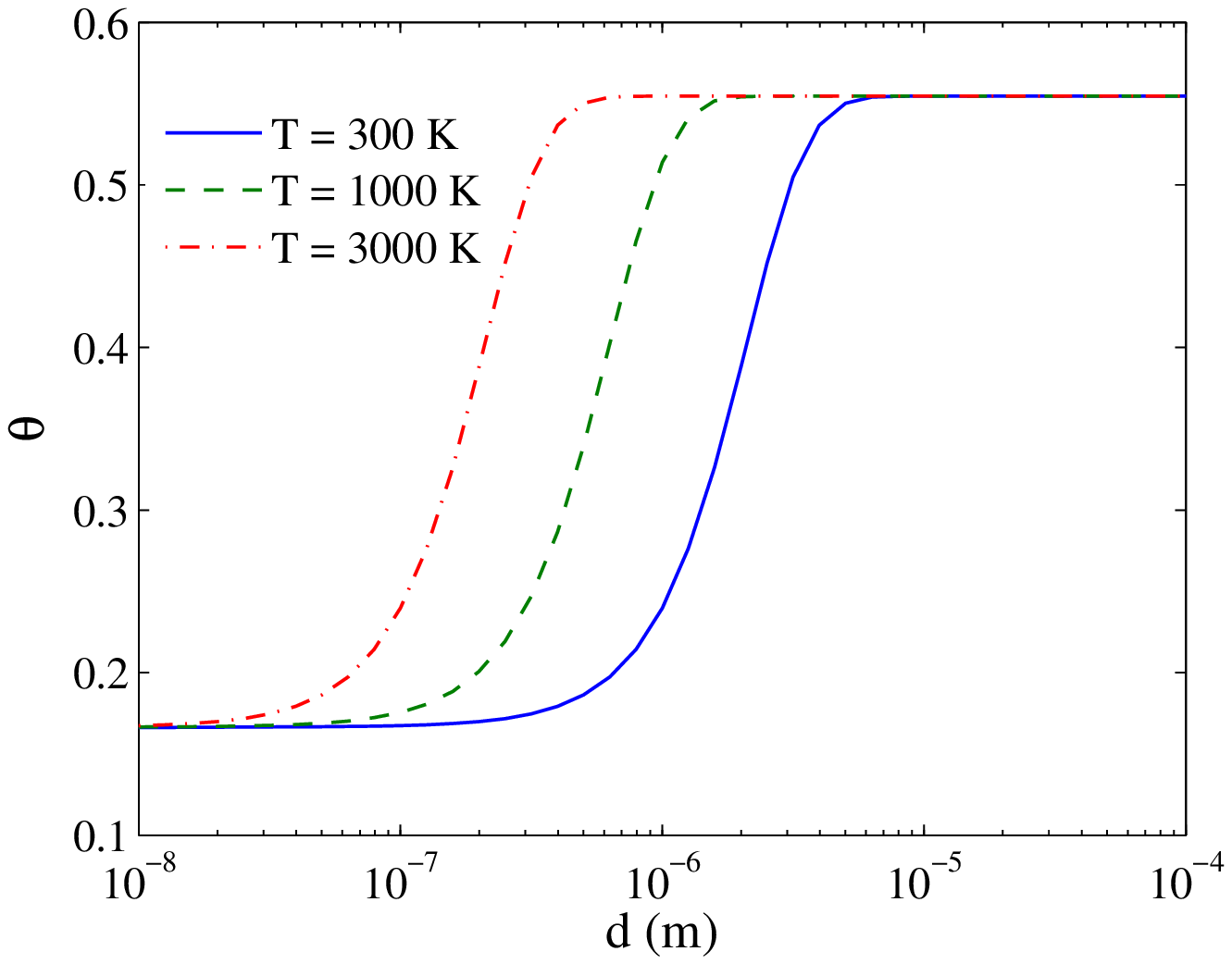} \caption{\label{f6} Dependence of $\theta$ on the distance between the spheres when $D=3$. }\end{figure}

\begin{figure}[h]
\epsfxsize=0.5\linewidth \epsffile{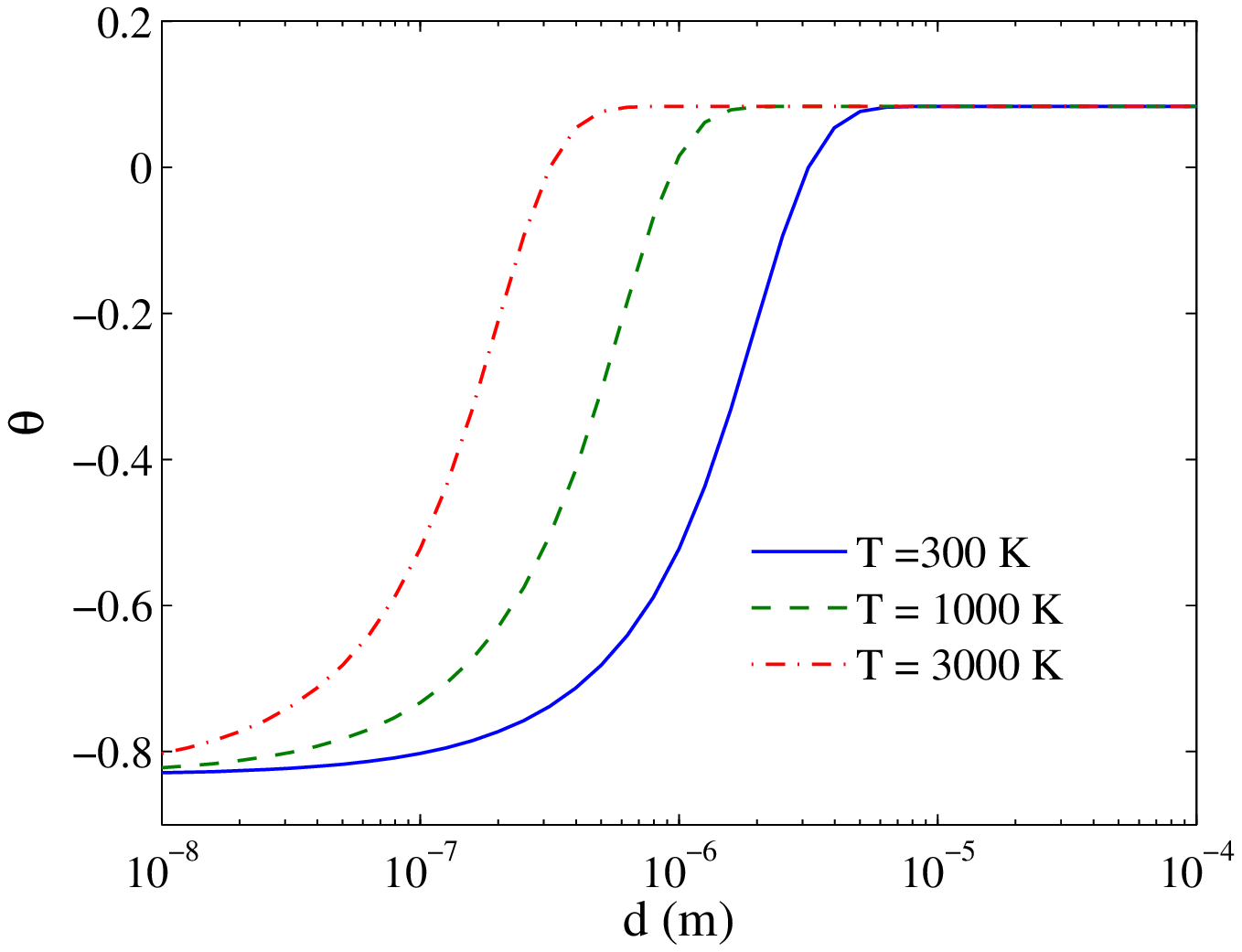} \caption{\label{f7} Dependence of $\theta$ on the distance between the spheres when $D=4$. }\end{figure}

\begin{figure}[h]
\epsfxsize=0.5\linewidth \epsffile{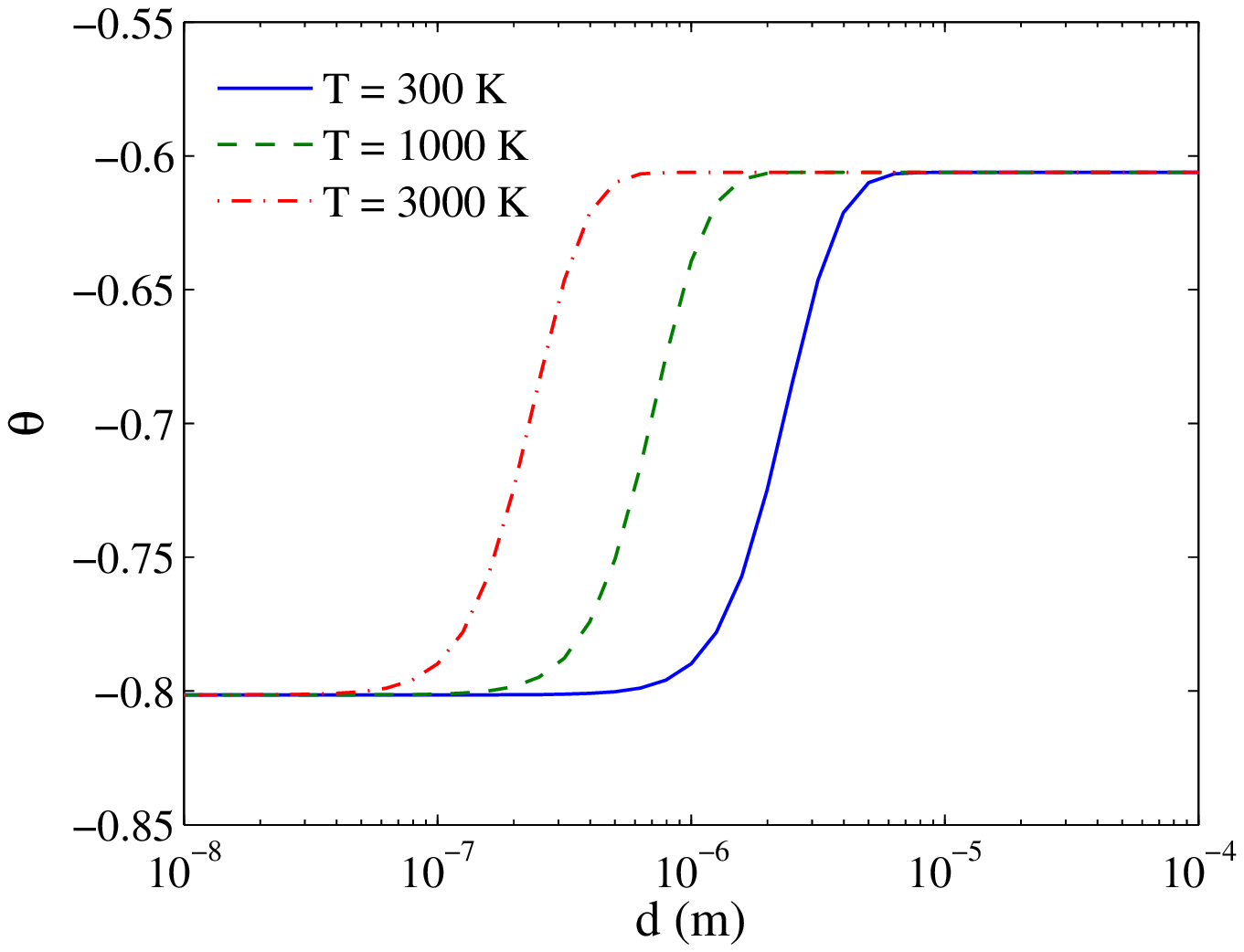} \caption{\label{f8} Dependence of $\theta$ on the distance between the spheres when $D=5$. }\end{figure}

\begin{figure}[h]
\epsfxsize=0.5\linewidth \epsffile{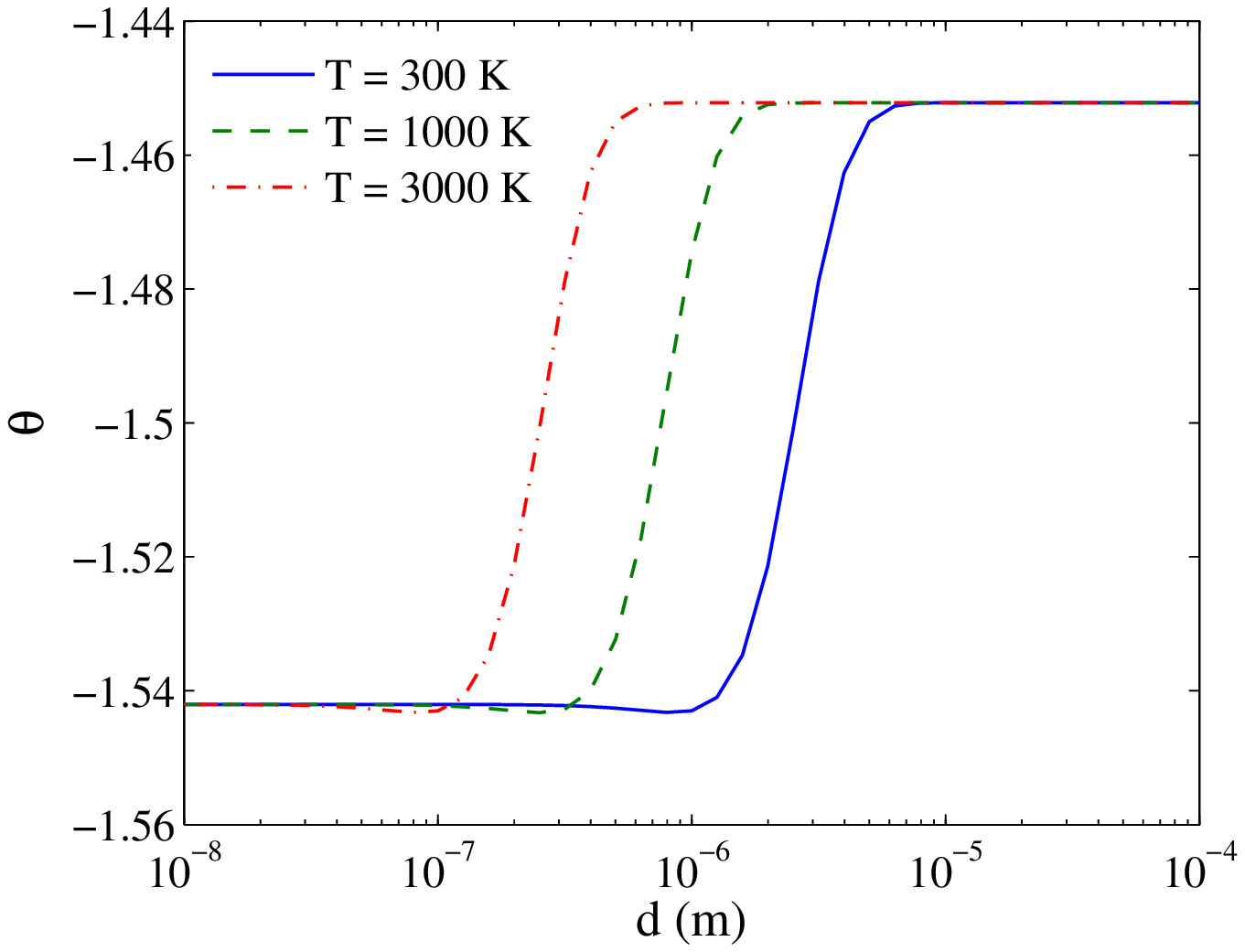} \caption{\label{f9} Dependence of $\theta$ on the distance between the spheres when $D=6$. }\end{figure}

Combining with the results from Section \ref{classical}, we find that the small separation leading term of the Casimir interaction force is 
\begin{equation}\label{eq6_5_2}
\begin{split}
F_{\text{Cas}}^0=&-\frac{(D-1)T\zeta(D)}{2^{\frac{3D-1}{2}} }\left(\frac{ R_1R_2}{R_1+R_2}\right)^{\frac{D-1}{2}}\frac{1}{d^{\frac{D+1}{2}}}\\&-\frac{ T  }{2^{D-3}    \Gamma\left(\frac{D-1}{2}\right)}\left(\frac{R_1R_2}{R_1+R_2}\right)^{\frac{D-1}{2}}\frac{1}{d^{\frac{D+1}{2}}}\sum_{k=1}^{\infty}\frac{1}{k^{\frac{D-1}{2}}}\sum_{p=1}^{\infty}
 \int_0^{\infty} dz \,\frac{ z^{D-2}}{\left(z^2+( \kappa_p d)^2\right)^{\frac{D-3}{4}}} \exp\left(
 - 2 k \sqrt{z^2+( \kappa_p d)^2} \right).
\end{split}\end{equation}
When $D=3$, the next-to--leading order term is 
\begin{equation}\label{eq6_5_3}
\begin{split}
F_{\text{Cas}}^1=&-\frac{T}{48d}+F_{\text{Cas}}^{\text{rem}, 1b};
\end{split}
\end{equation}when $D=4$,
\begin{equation}\label{eq6_5_4}
\begin{split}
F_{\text{Cas}}^1=&-\frac{3T\zeta(4)}{32\sqrt{2} }\left(\frac{ R_1R_2}{R_1+R_2}\right)^{\frac{3}{2}}\frac{1}{d^{\frac{5}{2}}} \left\{ -\frac{1}{4}\frac{d}{R_1+R_2}+ \frac{1}{12 } \left(\frac{d}{R_1}+\frac{d}{R_2}\right)\right\} +F_{\text{Cas}}^{\text{rem}, 1b};
\end{split}\end{equation}and when $D\geq 5$,
\begin{equation}\label{eq6_5_5}
\begin{split}
F_{\text{Cas}}^1=&-\frac{(D-1)T\zeta(D)}{2^{\frac{3D-1}{2}} }\left(\frac{ R_1R_2}{R_1+R_2}\right)^{\frac{D-1}{2}}\frac{1}{d^{\frac{D+1}{2}}}\\&\times\left\{ -\frac{D-3}{4}\frac{d}{R_1+R_2}+\left[\frac{D-3}{12 } -\frac{(D-3)(D-4)}{3(D-1) }\frac{\zeta(D-2)}{\zeta(D)}\right]\left(\frac{d}{R_1}+\frac{d}{R_2}\right) \right\}+F_{\text{Cas}}^{\text{rem}, 1a}+F_{\text{Cas}}^{\text{rem}, 1b}.
\end{split}
\end{equation}

Compare the leading term \eqref{eq6_5_2} to the proximity force approximation \eqref{eq5_23_7}, it is easy to check that under the change of variables
$$x=\sqrt{z^2+( \kappa_p d)^2},$$   \eqref{eq5_23_6} is equal to the sum of  the $p\neq 0$ terms in \eqref{eq5_23_7}. Together with   \eqref{eq5_26_2} and \eqref{eq5_22_1}, we find that the small separation leading term of the Casimir   interaction always agree with the proximity force approximation. This is true at any temperature. In fact, for $D=3$, we have proved this in the work \cite{54}. We would like to stress that this is a remarkable result since it holds at \emph{any} temperature.

Define
\begin{equation}
\theta=\frac{1}{\vep}\frac{F_{\text{Cas}}^1}{F_{\text{Cas}}^0}
\end{equation}to be the ratio of the next-to-leading order term to the leading order term divided by $\vep$. In Figs. \ref{f6}, \ref{f7}, \ref{f8} and \ref{f9}, we plot the dependence of $\theta$ on the distance between the spheres when $D=3, 4, 5$ and 6. We take the radius of both spheres to be $R=1$ mm.

Let us now obtain the explicit  small separation asymptotic expansions in different temperature regions. As in \cite{7}, we consider the following three regions:
\begin{enumerate}
\item[1.] High temperature: $1\ll dT \ll R_1\leq R_2T$,
\item[2.] Low temperature: $dT\ll R_1T\leq R_2T\ll 1$,
\item[3.] Medium temperature: $dT \ll 1\ll  R_1T\leq R_2T $.

\end{enumerate}

In the high temperature regime, it is easy to see that the dominating term is the classical term considered in Section \ref{classical} since \eqref{eq5_23_6}, \eqref{eq5_23_8} and \eqref{eq5_26_1} goes to zero exponentially fast when $dT\gg 1$. In considering the Casimir interaction at finite temperature, we have separated out the classical term which corresponds to $p=0$. The main reason is that the classical term can in fact be computed exactly. However, we have seen (eq. \eqref{eq5_22_1}) that by putting $p=0$ in the leading term \eqref{eq5_23_6}, we indeed obtain the leading term of the classical Casimir interaction force \eqref{eq5_26_2}. For $D\geq 5$, one can also check that putting $p=0$ in \eqref{eq5_23_8} and \eqref{eq5_26_1} will give exactly the same next-to-leading order term of the classical Casimir interaction force \eqref{eq5_26_3}.

Let us now consider the low temperature and medium temperature regions.
Using the inverse Mellin transform formula
\begin{equation}
e^{-z}=\frac{1}{2\pi i}\int_{c-i\infty}^{c+i\infty} dw\Gamma(w) z^{-w},
\end{equation}where $c$ is a positive constant, we find that
\begin{equation}\label{eq5_26_4}\begin{split}
F_{\text{Cas}}^{\text{rem}, 0}= & -\frac{ T  }{2^{D-3}    \Gamma\left(\frac{D-1}{2}\right)}\left(\frac{R_1R_2}{R_1+R_2}\right)^{\frac{D-1}{2}}
\frac{1}{d^{\frac{D+1}{2}}}\frac{1}{2\pi i}\int_{c-i\infty}^{c+i\infty} dw\Gamma(w)2^{-w}\zeta\left(w+\frac{D-1}{2}\right)    \sum_{p=1}^{\infty}
\int_0^{\infty}dz\frac{ z^{D-2}}{\left(z^2+( \kappa_p d )^2\right)^{\frac{w}{2}+\frac{D-3}{4}}}
 \\
 =& -\frac{ T  }{2^{D-2}    }\left(\frac{R_1R_2}{R_1+R_2}\right)^{\frac{D-1}{2}}
 \\&\times \frac{1}{2\pi i}\int_{c-i\infty}^{c+i\infty} dw\Gamma(w)2^{-w}\zeta\left(w+\frac{D-1}{2}\right)
\frac{\Gamma\left(\frac{w}{2}-\frac{D+1}{4}\right)}{\Gamma\left(\frac{w}{2}+\frac{D-3}{4}\right)} d^{-w}(2\pi T)^{\frac{D+1}{2}-w}\zeta\left(w-\frac{D+1}{2}\right).\end{split}\end{equation}
The pole of the integrand is at $\displaystyle w=\frac{D+3}{2}, \frac{D+1}{2}$ and $w=0, -1, -2,\ldots$. Evaluating the residues at these poles, we find that as $dT\ll 1$,
\begin{equation}\label{eq5_27_1}\begin{split}
F_{\text{Cas}}^{\text{rem}, 0}\sim  &-\frac{ T  }{2^{D-2}    }\left(\frac{R_1R_2}{R_1+R_2}\right)^{\frac{D-1}{2}} \left\{\frac{\zeta(D+1)}{2^{\frac{D+5}{2}}\sqrt{\pi}T d^{\frac{D+3}{2}}}\frac{\Gamma\left(\frac{D+3}{2}\right)}{\Gamma\left(\frac{D}{2}\right)}-\frac{(D-1)\zeta(D)}{2^{\frac{D+3}{2}}d^{\frac{D+1}{2}}}\right.\\
&\left.+\frac{\zeta\left(\frac{D-1}{2}\right)}
{\Gamma\left(\frac{D-3}{4}\right)}\frac{(2T)^{\frac{D+1}{2}}}{\sqrt{\pi}}\Gamma\left(\frac{D+3}{4}\right)\zeta\left(\frac{D+3}{2}\right)
-\frac{\zeta\left(\frac{D-3}{2}\right)}
{\Gamma\left(\frac{D-5}{4}\right)}\frac{2^{\frac{D+5}{2}}T^{\frac{D+3}{2}}}{\sqrt{\pi}}\Gamma\left(\frac{D+5}{4}\right)\zeta\left(\frac{D+5}{2}\right)d+\ldots\right\}.
\end{split}\end{equation}
When $D=3$, it is understood that
\begin{align}
\frac{\zeta\left(\frac{D-1}{2}\right)}
{\Gamma\left(\frac{D-3}{4}\right)}=\lim_{D\rightarrow 3} \frac{\zeta\left(\frac{D-1}{2}\right)}
{\Gamma\left(\frac{D-3}{4}\right)}=\frac{1}{2}.
\end{align}Similarly, when $D=5$,
\begin{align}
\frac{\zeta\left(\frac{D-3}{2}\right)}
{\Gamma\left(\frac{D-5}{4}\right)}=\lim_{D\rightarrow 5} \frac{\zeta\left(\frac{D-3}{2}\right)}
{\Gamma\left(\frac{D-5}{4}\right)}=\frac{1}{2}.
\end{align}
Combining with the leading term of the classical term \eqref{eq5_26_2}, we find that when $dT\ll 1$, the leading term of the Casimir interaction force is
\begin{equation}\label{eq5_27_2}\begin{split}
F_{\text{Cas}}^{ 0}\sim  &-\frac{\zeta(D+1)}{2^{\frac{3D+1}{2}}\sqrt{\pi}  }\frac{\Gamma\left(\frac{D+3}{2}\right)}{\Gamma\left(\frac{D}{2}\right)}\left(\frac{R_1R_2}{R_1+R_2}\right)^{\frac{D-1}{2}}
\frac{1}{d^{\frac{D+3}{2}}} -\frac{\zeta\left(\frac{D-1}{2}\right)}
{\Gamma\left(\frac{D-3}{4}\right)}\frac{ T^{\frac{D+3}{2}}}{2^{\frac{D-5}{2}}\sqrt{\pi}}\Gamma\left(\frac{D+3}{4}\right)\zeta\left(\frac{D+3}{2}\right)\left(\frac{R_1R_2}{R_1+R_2}\right)^{\frac{D-1}{2}}
\\&+\frac{\zeta\left(\frac{D-3}{2}\right)}
{\Gamma\left(\frac{D-5}{4}\right)}\frac{T^{\frac{D+5}{2}}d}{2^{\frac{D-9}{2}}\sqrt{\pi}}\Gamma\left(\frac{D+5}{4}\right)\zeta\left(\frac{D+5}{2}\right)
\left(\frac{R_1R_2}{R_1+R_2}\right)^{\frac{D-1}{2}}+\ldots.
\end{split}\end{equation}
The first term
\begin{align}F_{\text{Cas}}^{T=0, 0}=-\frac{\zeta(D+1)}{2^{\frac{3D+1}{2}}\sqrt{\pi}  }\frac{\Gamma\left(\frac{D+3}{2}\right)}{\Gamma\left(\frac{D}{2}\right)}\left(\frac{R_1R_2}{R_1+R_2}\right)^{\frac{D-1}{2}}
\frac{1}{d^{\frac{D+3}{2}}} \end{align} is the zero temperature leading term \eqref{eq5_22_2}. The next two terms in \eqref{eq5_27_2} comes from finite temperature contributions. Namely, the leading terms of the temperature correction is
\begin{equation}\label{eq5_30_5}\begin{split}
\Delta_TF_{\text{Cas}}^{ 0}\sim  & -\frac{\zeta\left(\frac{D-1}{2}\right)}
{\Gamma\left(\frac{D-3}{4}\right)}\frac{ T^{\frac{D+3}{2}}}{2^{\frac{D-5}{2}}\sqrt{\pi}}\Gamma\left(\frac{D+3}{4}\right)\zeta\left(\frac{D+3}{2}\right)\left(\frac{R_1R_2}{R_1+R_2}\right)^{\frac{D-1}{2}}
\\&+\frac{\zeta\left(\frac{D-3}{2}\right)}
{\Gamma\left(\frac{D-5}{4}\right)}\frac{T^{\frac{D+5}{2}}d}{2^{\frac{D-9}{2}}\sqrt{\pi}}\Gamma\left(\frac{D+5}{4}\right)\zeta\left(\frac{D+5}{2}\right)
\left(\frac{R_1R_2}{R_1+R_2}\right)^{\frac{D-1}{2}}+\ldots.
\end{split}\end{equation}
They become significant when $dT\sim 1$.

Using the same method, we can compute the asymptotic expansion of the next-to-leading order term $F_{\text{Cas}}^{\text{rem}, 1}$ when $dT\ll 1$. When $D=3,4,5$, $F_{\text{Cas}}^{\text{rem}, 1a}=0$. For $D\geq 6$,
\begin{equation}\label{eq5_27_4}\begin{split}
F_{\text{Cas}}^{\text{rem}, 1a}= &  \frac{(D-5)}{3}\frac{ 1  }{2^{\frac{3D+1}{2}} \sqrt{\pi}  }\left(\frac{R_1R_2}{R_1+R_2}\right)^{\frac{D-3}{2}}  \frac{\Gamma\left(\frac{D+1}{2}\right)}{\Gamma\left(\frac{D-2}{2}\right)}
\frac{\zeta(D-1)}{  d^{\frac{D+1}{2}}}\\
&-\frac{T}{2^{\frac{3D-1}{2}}}\left(\frac{R_1R_2}{R_1+R_2}\right)^{\frac{D-3}{2}}\frac{\zeta(D-2)}{d^{\frac{D-1}{2}}}\frac{(D-3)(D-5)}{3}\\
&+\frac{(D-5)}{3}\frac{ T^{\frac{D+1}{2}}   }{2^{\frac{D-1}{2}}   \sqrt{\pi} }\left(\frac{R_1R_2}{R_1+R_2}\right)^{\frac{D-3}{2}}\frac{\Gamma\left(\frac{D+1}{4}\right)}{\Gamma\left(\frac{D-5}{4}\right)}
 \zeta\left(\frac{D-3}{2}\right)\zeta\left(\frac{D+1}{2}\right)\\
 &-\frac{(D-5)}{3}\frac{ T^{\frac{D+3}{2}}  d }{2^{\frac{D-5}{2}}   \sqrt{\pi} }\left(\frac{R_1R_2}{R_1+R_2}\right)^{\frac{D-3}{2}}\frac{\Gamma\left(\frac{D+3}{4}\right)}{\Gamma\left(\frac{D-7}{4}\right)}
 \zeta\left(\frac{D-5}{2}\right)\zeta\left(\frac{D+3}{2}\right)+\ldots.
\end{split}\end{equation}
When $D=3$,
\begin{equation}\label{eq5_29_2}\begin{split}
F_{\text{Cas}}^{\text{rem}, 1b}= & -\frac{R_1R_2\zeta(4)}{16\pi (R_1+R_2)d^3}\left[\frac{1}{3}\left(\frac{d}{R_1}+\frac{1}{R_2}\right)-\frac{d}{R_1+R_2}\right]+\frac{T}{48d}\\&+\frac{\pi T^2}{12}\frac{R_1R_2}{R_1+R_2}
\left[\frac{1}{3}\left(\frac{\pi^2}{6}-1\right)\left(\frac{1}{R_1}+\frac{1}{R_2}\right)-\frac{\pi^2}{6}\frac{1}{R_1+R_2}\right]-\frac{dT^3R_1R_2\zeta(3)}{(R_1+R_2)}\left[\frac{1}{4}
\left(\frac{1}{R_1}+\frac{1}{R_2}\right)-\frac{1}{R_1+R_2}\right].
\end{split}\end{equation}
For $D\geq 4$,
\begin{equation}\label{eq5_27_5}\begin{split}
&F_{\text{Cas}}^{\text{rem}, 1b}\\
=&\frac{ 1  }{2^{\frac{3D+1}{2}}\sqrt{\pi}      }\left(\frac{R_1R_2}{R_1+R_2}\right)^{\frac{D-1}{2}}
\frac{ \Gamma\left(\frac{D+1}{2}\right) }{\Gamma\left(\frac{D}{2}\right)}\frac{\zeta(D+1)}{d^{\frac{D+3}{2}}}\left\{
\frac{(D^2-1)}{8} \frac{d}{R_1+R_2} -\left[\frac{D^2-1}{24}-  \frac{\zeta(D-1)}{\zeta(D+1) }
\frac{(D+3)(D-2)}{6D}\right]\left(\frac{d}{R_1}+\frac{d}{R_2}\right)\right\}\\
&+\frac{(D-1)T\zeta(D)}{2^{\frac{3D-1}{2}} }\left(\frac{ R_1R_2}{R_1+R_2}\right)^{\frac{D-1}{2}}\frac{1}{d^{\frac{D+1}{2}}}
\left\{-\frac{D-3}{4}\frac{d}{R_1+R_2}+\left[\frac{D-3}{12 } -\frac{(D-3)}{3(D-1) }\frac{\zeta(D-2)}{\zeta(D)}\right]\left(\frac{d}{R_1}+\frac{d}{R_2}\right)\right\}\\
&-\frac{ T^{\frac{D+1}{2}}  }{2^{\frac{D+1}{2}}\sqrt{\pi}      }\left(\frac{R_1R_2}{R_1+R_2}\right)^{\frac{D-1}{2}}
\frac{ \Gamma\left(\frac{D+1}{4}\right) }{\Gamma\left(\frac{D-1}{4}\right)} \zeta\left(\frac{D+1}{2}\right)^2\left\{
\frac{(D^2-1) }{4} \frac{1}{R_1+R_2}-\left[
\frac{(D^2-1) }{12}-\frac{\zeta\left(\frac{D-3}{2}\right)}{\zeta\left(\frac{D+1}{2}\right)}\frac{D-5}{3} \right]\left(\frac{1}{R_1}+\frac{1}{R_2}\right)\right\} \\
&+\frac{ T^{\frac{D+3}{2}}d  }{2^{\frac{D-3}{2}}\sqrt{\pi}      }\left(\frac{R_1R_2}{R_1+R_2}\right)^{\frac{D-1}{2}}
\frac{ \Gamma\left(\frac{D+3}{4}\right) }{\Gamma\left(\frac{D-3}{4}\right)} \zeta\left(\frac{D+3}{2}\right)\zeta\left(\frac{D-1}{2}\right)\left\{
\frac{(D-1)(D+5) }{4} \frac{1}{R_1+R_2}\right.\\&\left.\hspace{6cm}-\left[
\frac{(D-1)(D+5) }{12}-\frac{\zeta\left(\frac{D-5}{2}\right)}{\zeta\left(\frac{D-1}{2}\right)}\frac{(D-7)(D+3)}{3(D-3)} \right]\left(\frac{1}{R_1}+\frac{1}{R_2}\right)\right\}.
\end{split}\end{equation}

From \eqref{eq6_5_3} and \eqref{eq5_29_2}, we find that when $D=3$, the next-to-leading order term of the small separation asymptotic expansion of the Casimir interaction force is
\begin{equation}\label{eq5_29_1}\begin{split}
F_{\text{Cas}}^{1}\sim &-\frac{R_1R_2\zeta(4)}{16\pi (R_1+R_2)d^3}\left[\frac{1}{3}\left(\frac{d}{R_1}+\frac{1}{R_2}\right)-\frac{d}{R_1+R_2}\right]+\frac{\pi T^2}{12}\frac{R_1R_2}{R_1+R_2}
\left[\frac{1}{3}\left(\frac{\pi^2}{6}-1\right)\left(\frac{1}{R_1}+\frac{1}{R_2}\right)-\frac{\pi^2}{6}\frac{1}{R_1+R_2}\right]\\
&-\frac{dT^3R_1R_2\zeta(3)}{(R_1+R_2)}\left[\frac{1}{4}\left(\frac{1}{R_1}+\frac{1}{R_2}\right)-\frac{1}{R_1+R_2}\right].
\end{split}
\end{equation}The first term is the zero temperature part, whereas the remaining terms are temperature corrections.
Combining with \eqref{eq6_5_2}, we find that when $d\ll R_1\leq R_2$, the zero temperature Casimir interaction force behaves as
\begin{align}\label{eq6_2_1}
F_{\text{Cas}}^{T=0}\sim -\frac{R_1R_2\zeta(4)}{8\pi (R_1+R_2)d^3}\left\{1-\frac{1}{2}\frac{d}{R_1+R_2}+\frac{1}{6}\left(\frac{d}{R_1}+\frac{d}{R_2}\right)+\ldots\right\}.
\end{align}
This is the small separation asymptotic expansion in the low temperature regime.

For the temperature correction, we have
\begin{equation}\begin{split}
\Delta_TF_{\text{Cas}}\sim &-\frac{\zeta(3)T^3}{2}\frac{R_1R_2}{R_1+R_2}+\frac{\pi^3T^4d}{45}\frac{R_1R_2}{R_1+R_2}+\ldots+\frac{\pi T^2}{12}\frac{R_1R_2}{R_1+R_2}
\left[\frac{1}{3}\left(\frac{\pi^2}{6}-1\right)\left(\frac{1}{R_1}+\frac{1}{R_2}\right)-\frac{\pi^2}{6}\frac{1}{R_1+R_2}\right]
\\&-\frac{dT^3R_1R_2\zeta(3)}{(R_1+R_2)}\left[\frac{1}{4}\left(\frac{1}{R_1}+\frac{1}{R_2}\right)-\frac{1}{R_1+R_2}\right]+\ldots
\end{split}\end{equation}This, together with \eqref{eq6_2_1}, give the small separation asymptotic expansion in the medium temperature regime.

 When $D=4$, \eqref{eq6_5_2}, \eqref{eq6_5_4} and \eqref{eq5_27_5}  give
\begin{equation}\label{eq6_5_1}
\begin{split}
F_{\text{Cas}}^{T=0}=&-\frac{15\zeta(5)}{512\sqrt{2}}\left(\frac{R_1R_2}{R_1+R_2}\right)^{\frac{3}{2}}\frac{1}{d^{\frac{7}{2}}}
\left\{1-\frac{3}{4}\frac{d}{R_1+R_2}+\left[\frac{1}{4}-\frac{7}{30}\frac{\zeta(3)}{\zeta(5)}\right]\left(\frac{d}{R_1}+\frac{d}{R_2}\right)+\ldots\right\},
\end{split}
\end{equation}
\begin{equation}
\begin{split}
\Delta_TF_{\text{Cas}}=&-\frac{\pi^2T}{576\sqrt{2}}\left(\frac{R_1R_2}{R_1+R_2}\right)^{\frac{1}{2}}\frac{1}{d^{\frac{3}{2}}}
-\sqrt{2}\frac{\Gamma\left(\frac{7}{4}\right)}{\Gamma\left(\frac{1}{4}\right)}\frac{T^{\frac{7}{2}}}{\sqrt{\pi}}\zeta\left(\frac{3}{2}\right)\zeta\left(\frac{7}{2}\right)
\left(\frac{R_1R_2}{R_1+R_2}\right)^{\frac{3}{2}}\\&+4\sqrt{2}\frac{\Gamma\left(\frac{9}{4}\right)}{\Gamma\left(-\frac{1}{4}\right)}\frac{T^{\frac{9}{2}}d}{\sqrt{\pi}}\zeta\left(\frac{1}{2}\right)\zeta\left(\frac{9}{2}\right)
\left(\frac{R_1R_2}{R_1+R_2}\right)^{\frac{3}{2}}\\
&-\frac{1}{4\sqrt{2}}\frac{\Gamma\left(\frac{5}{4}\right)}{\Gamma\left(\frac{3}{4}\right)}\frac{T^{\frac{5}{2}}}{\sqrt{\pi}}\zeta\left(\frac{5}{2}\right)^2
\left(\frac{R_1R_2}{R_1+R_2}\right)^{\frac{3}{2}}\left\{\frac{15}{4}\frac{1}{R_1+R_2}-\left[\frac{5}{4}+\frac{1}{3}\frac{\zeta\left(\frac{1}{2}\right)}{\zeta\left(\frac{5}{2}\right)}
\right]\left(\frac{1}{R_1}+\frac{1}{R_2}\right)\right\}\\
&+\frac{1}{\sqrt{2}}\frac{\Gamma\left(\frac{7}{4}\right)}{\Gamma\left(\frac{1}{4}\right)}\frac{T^{\frac{7}{2}}d}{\sqrt{\pi}}\zeta\left(\frac{7}{2}\right)\zeta\left(\frac{3}{2}\right)
\left(\frac{R_1R_2}{R_1+R_2}\right)^{\frac{3}{2}}\left\{\frac{27}{4}\frac{1}{R_1+R_2}-\left[\frac{9}{4}+7\frac{\zeta\left(-\frac{1}{2}\right)}{\zeta\left(\frac{3}{2}\right)}
\right]\left(\frac{1}{R_1}+\frac{1}{R_2}\right)\right\}+\ldots
\end{split}
\end{equation}

When $D\geq 5$, \eqref{eq6_5_2}, \eqref{eq6_5_5}, \eqref{eq5_27_4}, \eqref{eq5_27_5} give  
\begin{equation}\label{eq5_29_4}\begin{split}
F_{\text{Cas}}^{T=0}=&-\frac{\zeta(D+1)}{2^{\frac{3D+1}{2}}\sqrt{\pi}  }\frac{\Gamma\left(\frac{D+3}{2}\right)}{\Gamma\left(\frac{D}{2}\right)}\left(\frac{R_1R_2}{R_1+R_2}\right)^{\frac{D-1}{2}}
\frac{1}{d^{\frac{D+3}{2}}}\left\{1-\frac{D-1}{4}\frac{d}{R_1+R_2}\right.\\&\left.\hspace{5cm}
+\left[\frac{D-1}{12}-\frac{\zeta(D-1)}{\zeta(D+1)}\frac{(D-1)(D-2)(D-3)}{3D(D+1)}\right]\left(\frac{d}{R_1}+\frac{d}{R_2}\right)+\ldots\right\}.
\end{split}\end{equation}
This gives the small separation asymptotic expansion in the low temperature regime.

For the temperature correction, we have
\begin{equation}\label{eq5_29_5}
\begin{split}
\Delta_TF_{\text{Cas}}=& -\frac{\zeta\left(\frac{D-1}{2}\right)}
{\Gamma\left(\frac{D-3}{4}\right)}\frac{ T^{\frac{D+3}{2}}}{2^{\frac{D-5}{2}}\sqrt{\pi}}\Gamma\left(\frac{D+3}{4}\right)\zeta\left(\frac{D+3}{2}\right)\left(\frac{R_1R_2}{R_1+R_2}\right)^{\frac{D-1}{2}}
\\&+\frac{\zeta\left(\frac{D-3}{2}\right)}
{\Gamma\left(\frac{D-5}{4}\right)}\frac{T^{\frac{D+5}{2}}}{2^{\frac{D-9}{2}}\sqrt{\pi}}\Gamma\left(\frac{D+5}{4}\right)\zeta\left(\frac{D+5}{2}\right)
\left(\frac{R_1R_2}{R_1+R_2}\right)^{\frac{D-1}{2}}d+\ldots\\
&-\frac{ T^{\frac{D+1}{2}}  }{2^{\frac{D+1}{2}}\sqrt{\pi}      }\left(\frac{R_1R_2}{R_1+R_2}\right)^{\frac{D-1}{2}}
\frac{ \Gamma\left(\frac{D+1}{4}\right) }{\Gamma\left(\frac{D-1}{4}\right)} \zeta\left(\frac{D+1}{2}\right)^2\left\{
\frac{(D^2-1) }{4} \frac{1}{R_1+R_2}\right.\\&\left.\hspace{3cm}-\left[
\frac{(D^2-1) }{12}+\frac{\zeta\left(\frac{D-3}{2}\right)}{\zeta\left(\frac{D+1}{2}\right)}\frac{(D-5)(D-7)}{6} \right]\left(\frac{1}{R_1}+\frac{1}{R_2}\right)\right\} \\
&+\frac{ T^{\frac{D+3}{2}}d  }{2^{\frac{D-3}{2}}\sqrt{\pi}      }\left(\frac{R_1R_2}{R_1+R_2}\right)^{\frac{D-1}{2}}
\frac{ \Gamma\left(\frac{D+3}{4}\right) }{\Gamma\left(\frac{D-3}{4}\right)} \zeta\left(\frac{D+3}{2}\right)\zeta\left(\frac{D-1}{2}\right)\left\{
\frac{(D-1)(D+5) }{4} \frac{1}{R_1+R_2}\right.\\&\left.\hspace{2cm}-\left[
\frac{(D-1)(D+5) }{12}+\frac{\zeta\left(\frac{D-5}{2}\right)}{\zeta\left(\frac{D-1}{2}\right)}\frac{(D-1)(D-7)(D-9)}{6(D-3)} \right]\left(\frac{1}{R_1}+\frac{1}{R_2}\right)\right\}+\ldots.
\end{split}
\end{equation} \eqref{eq5_29_4} together with \eqref{eq5_29_5} give the small separation asymptotic expansion of the Casimir interaction force in the medium temperature regime.

One can check that the zero temperature asymptotic expansions \eqref{eq6_2_1}, \eqref{eq6_5_1} and \eqref{eq5_29_4} agree with the result we derived in \cite{1}.

From the graphs Figs. \ref{f6}, \ref{f7}, \ref{f8} and \ref{f9}, we find that when $dT\ll 1$, $\theta$ approaches a limiting value given by the zero temperature limit
 \begin{equation}
 \theta^{T=0}=\left\{\begin{aligned} &-\frac{1}{2} +\frac{1}{6 a_1a_2},\hspace{1cm}& D=3,\\
& -\frac{3}{4} +\left[\frac{1}{4}-\frac{7}{30}\frac{\zeta(3)}{\zeta(5)}\right]\frac{1}{a_1a_2},\hspace{1cm} & D=4,\\
&-\frac{D-1}{4} 
+\left[\frac{D-1}{12}-\frac{\zeta(D-1)}{\zeta(D+1)}\frac{(D-1)(D-2)(D-3)}{3D(D+1)}\right]\frac{1}{a_1a_2},\hspace{0.5cm} & D\geq 5;
 \end{aligned}\right.
 \end{equation}whereas as $dT\gg 1$, $\theta$ approaches a limiting value given by the high temperature limit
 \begin{equation}
 \theta^{\text{classical}}=\left\{\begin{aligned} &\frac{1}{6 \zeta(3)a_1a_2},\hspace{1cm}& D=3,\\
& -\frac{D-3}{4} +\left[\frac{D-3}{12 } -\frac{(D-3)(D-4)}{3(D-1) }\frac{\zeta(D-2)}{\zeta(D)}\right]\frac{1}{a_1a_2},\hspace{0.5cm} & D\geq 4.
 \end{aligned}\right.
 \end{equation}

\section{Conclusion}

In this work, we have considered the finite temperature Casimir interaction between two spheres subject to Dirichlet boundary conditions in $(D+1)$-dimensional Minkowski spacetime. Starting with the TGTG formula for the zero temperature Casimir interaction energy we derived in \cite{1}, we use Matsubara formalism to obtain the finite temperature Casimir free interaction energy. The term corresponding to zero Matsubara frequency is singled out. It gives the high temperature limit of the Casimir interaction and is known as the classical term. This term can be computed exactly by some similarity transformations of matrices. We then use Abel-Plana summation formula to obtain an alternative expression that can be used to deduce the small separation asymptotic expansions.

For the remaining part of the Casimir interaction, we use our results in the zero temperature case \cite{1} together with Matsubara formalism to derive its small separation leading and next-to-leading order terms which are valid at any temperature. Combining with the classical term, we find that the leading order term agrees with the proximity force approximation at any temperature. This is a remarkable result. An inverse Mellin transform is then used to compute the analytic expression of the leading and next-to-leading order terms in the low and medium temperature regions. In the low temperature region, the dominating term is the zero temperature term. In the medium temperature region, we have to take into account the contribution from the finite temperature corrections.

\begin{acknowledgments}\noindent
  This work is supported by the Ministry of Higher Education of Malaysia  under   FRGS grant FRGS/1/2013/ST02/UNIM/02/2.
\end{acknowledgments}

\end{document}